\documentclass[aps,twocolumn,nofootinbib,floatfix,superscriptaddress]{revtex4-2}

\usepackage[english]{babel}
\usepackage[utf8]{inputenc}
\usepackage[T1]{fontenc} 
\usepackage[normalem]{ulem}
\usepackage{amsmath,amssymb,amsfonts,physics,bm,graphicx,xcolor,mathtools,dsfont}
\usepackage[colorlinks=true,linkcolor={blue!35!green},citecolor={blue!70!black},urlcolor={blue!60!black}]{hyperref}
\usepackage[nameinlink]{cleveref}
\usepackage{mathrsfs}
\usepackage{xcolor}

\usepackage{upgreek}

\newcommand{\blue}[1]{\textcolor{blue}{#1}}

\let\oldbibitem\bibitem 
\renewcommand{\bibitem}{
    \renewcommand{\doi}[1]{\texttt{\href{https://doi.org/##1}{doi:##1}}} 
    \let\bibitem\oldbibitem 
    \oldbibitem 
}

\begin{document}

\title{Pb transport on Si(111)-($7\times7$) following tip induced imbalance}

\author{Paul Philip Schmidt\normalfont\textsuperscript{$\dagger$}}
\email{paulschm@uni-potsdam.de}
\affiliation{University of Potsdam, Institute of Physics and Astronomy, Karl--Liebknecht--Str. 24--25, 14476 Potsdam, Germany}
\author{Felix Hartmann\normalfont\textsuperscript{$\dagger$}}
\affiliation{University of Potsdam, Institute of Physics and Astronomy, Karl--Liebknecht--Str. 24--25, 14476 Potsdam, Germany}
\author{Lea Faber}
\affiliation{University of Potsdam, Institute of Physics and Astronomy, Karl--Liebknecht--Str. 24--25, 14476 Potsdam, Germany}
\author{Ralf Metzler}
\affiliation{University of Potsdam, Institute of Physics and Astronomy, Karl--Liebknecht--Str. 24--25, 14476 Potsdam, Germany}
\affiliation{Asia Pacific Center for Theoretical Physics, Pohang 37673, Republic of Korea}
\author{Janet Anders}
\affiliation{University of Potsdam, Institute of Physics and Astronomy, Karl--Liebknecht--Str. 24--25, 14476 Potsdam, Germany}
\affiliation{Department of Physics and Astronomy, University of Exeter, Stocker Road, Exeter EX4 4QL, UK}
\author{Regina Hoffmann-Vogel}
\affiliation{University of Potsdam, Institute of Physics and Astronomy, Karl--Liebknecht--Str. 24--25, 14476 Potsdam, Germany}

\def\thefootnote{$\dagger$}\footnotetext{These authors contributed equally to this work}

\begin{abstract}
    Pb on Si(111)-($7\times7$) shows surprising nucleation and mass transport dynamics at odds with standard theories. To create local imbalances on stable Pb islands we use the tip of a scanning force microscope. We enforce a short, local contact between the island and our tip. The subsequent island height growth and the local contact potential difference are studied via scanning force microscopy and Kelvin probe force microscopy. Though the island has a large volume increase after the contact, we observe that its surrounding wetting layer shows the same Pb density decrease as the global wetting layer. This indicates a collective density thinning of the wetting layer.
\end{abstract}

\maketitle

The diffusion behavior of metals on semiconductors is not only of fundamental physical importance, but also relevant for applications, e.g. in the semiconductor industry. In this manuscript we focus on Pb/Si(111)-($7\times7$) that we study by scanning force microscopy (SFM) and Kelvin probe force microscopy (KPFM), which allows us to access the local contact potential difference (LCPD). 
This is used to study the mass transport in the Pb wetting layer (WL).
The mass transport is one striking feature of the Pb/Si(111)-($7\times7$) system, which does not follow the conventional theory of random walk diffusion, as expected from the theory of nucleation by Stranski-Krastanov~\cite{Venables84p1,Baskaran_2012,Kesaria_2009,Prieto_2017,BAUER_1958}. One expects that individual clusters form and slowly evolve into islands. However, experimentally no cluster formation is observed~\cite{hershberger14p1,hupalo01p1,Hong_2007}. Instead, mature crystalline islands are found as soon as a critical coverage density is exceeded in the wetting layer (WL), i.e. the formation happens ``ultrafast''~\cite{jaroch2019,hershberger14p1,hupalo01p1,Ganz_1991,Weitering_1992,Wang_2008}. Together with laser induced thermal desorption (LITD) experiments studying the dynamics of the WL~\cite{Huang2012,hupalo07p1,Man08p1,man2013}, this points towards a collective movement of Pb atoms in the WL.

After the initial nucleation, Pb islands have two growth modes. They can grow radially (parallel to the Si surface) or in height (perpendicular to the Si surface)~\cite{kuntov2008}. The latter is known to energetically favor the growth of rings at the edge of the island, which later close towards the island center~\cite{Li09p1}. This follows from azimuthal diffusion on the island edge, which is orders of magnitude faster than radial diffusion towards the island center~\cite{kuntova2007}.
The growth mode is strongly related to the initial substrate temperature during evaporation. After evaporation at $120$~K, only small islands with flat tops and steep edges are formed \cite{hershberger14p1,hupalo01p1}. 
These then grow radially, but only rarely in height. 
Therefore, ring formation is almost never observed for low substrate temperatures. 
Through subsequent annealing to $\sim 220\,$K small islands disappear and the system forms only a few large, relatively stable islands. 
After evaporation at room temperature, relatively large islands are formed directly after evaporation. 
These islands are metastable as they show ring formation and growth in height~\cite{Kuntova2006}. 
Under either of both conditions, the initial ring formation cannot be observed. 

Recent studies have focused on the electron-phonon coupling~\cite{Tajik_2023}, the relaxation dynamics of hot carries~\cite{Kratzer_2022}, and the formation of charge- and spin-waves~\cite{Tresca_2023,Vandelli_2024} in thin Pb films on Si(111).
It remains an open question to understand the exact dynamics of the Pb atoms in the Pb/Si(111)-($7\times7$) system and how the WL contributes to it.
Are individual ``free'' Pb atoms the cause for the ultrafast dynamics? Or does the WL move in a collective fashion~\cite{Huang2012}?
In order to study this question, researchers have stimulated ring growth artificially. One approach was to evaporate additional Pb and to image the surface before and after the additional evaporation~\cite{hershberger14p1}. 
Moreover, island growth has been stimulated using bias pulses~\cite{SLi06p01} or by charging the islands in STM measurements~\cite{Han04p1}.

\begin{figure*}
    \includegraphics[width=\textwidth]{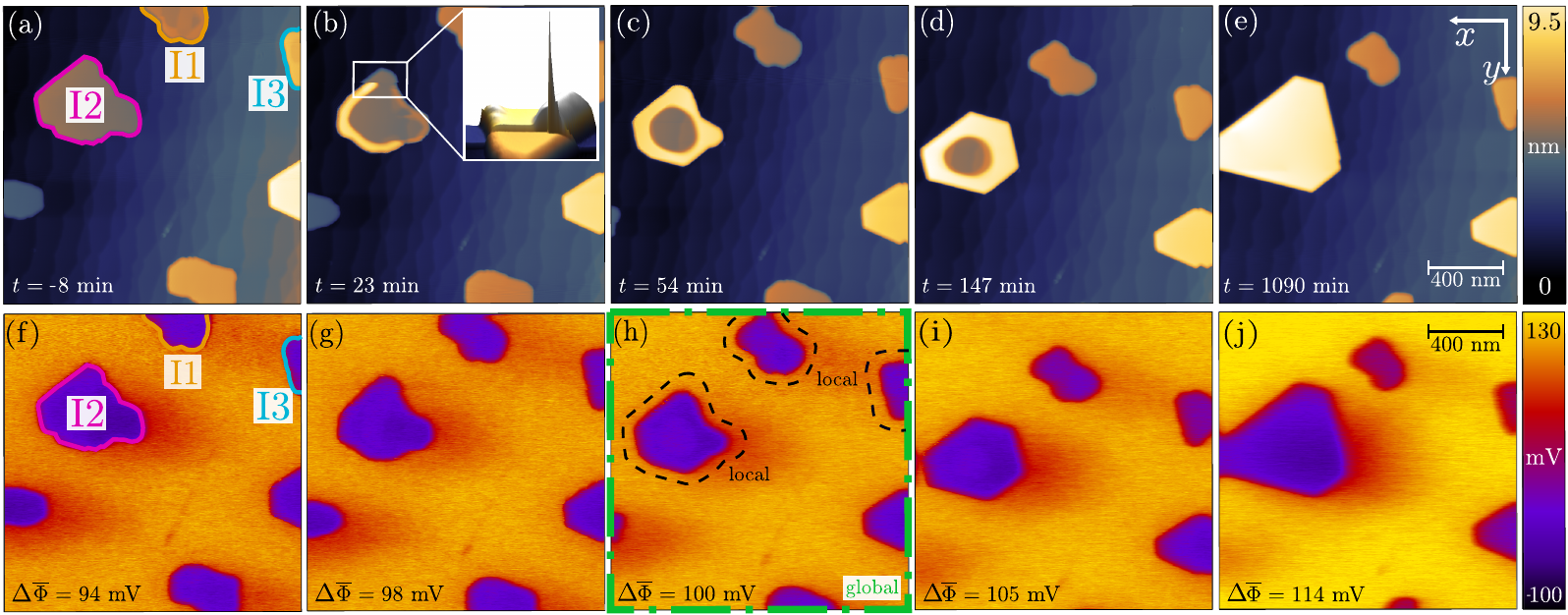}
    \caption{SFM images of 5 ML Pb on Si(111) after an annealing time of $34$~h at $300$~K. In (a)-(e) the measured SFM-topography, and in (f)-(j) the simultaneously measured local contact potential difference (LCPD) is shown.
    The first pixel after the tip-island contact sets the measurement time to $t = 0$~min, i.e. $8$~min after panel (a) is completely scanned.
    The time specifications refer to the fully scanned image, where each panel scan takes $31$~min.   
    (a) The islands of interest are labeled I1 (orange), I2 (magenta), and I3 (blue). 
    The inset in (b) shows a side-view image where the spike marks the tip-island contact.
    In (b) the ring formation is observed immediately after the tip-island contact.
    (c) After $39$~min (time between contact and fully scanned I2 in (c)) the ring is fully closed and the right-hand side of island I2 is reshaped. 
    (d)-(e) The island reshapes into a hexagonal shape and the ring closes towards the island center. After $1090$~min, the island surface is completely flat and in a new stable configuration. 
    (f)-(j) The local changes of the LCPD $\Delta\Phi$ (in a small neighborhood, see dashed black line in (h)) around island I2 are the same as observed globally within the image boundaries (see dashed-dotted green line in (h)) throughout the measurement, i.e. no local depletion of Pb is indicated.
    The mean LCPD $\Delta\overline{\Phi}$ of the WL increases globally from $94$~mV to $114$~mV (indicated by an increase in yellow color). In contrast, the $\Delta\overline{\Phi}$ of the islands remain constant (purple color).
    Measurement settings are $\Delta f = -37$~Hz, with an oscillation amplitude $A_{\mathrm{osc}} = 6.4$~nm.
    The SFM images (a)-(e) are leveled such that the manipulated island I2 appears at constant height. 
    In Appendix~\ref{app:heightprofile} the data is instead levelled with respect to the Si substrate.
    Further, the color scale in the LCPD images (f)-(j) is chosen such that the changes in the WL are emphasized. For a color scale emphasizing the QSE on the Pb islands see Appendix~\ref{app:QSE}.
    }
    \label{fig:kollision}
\end{figure*}
In this manuscript, we present a new experiment to locally generate imbalances by manipulating stable Pb islands. We cause a short contact between our tip and an initially stable Pb island in our SFM. 
This triggers the island to grow in height, showing rapid formation of a ring. 
This process is different from initial island growth and provides valuable information about the Pb flux from the WL to the island as equilibrium is reached.
On a much slower timescale the ring closes towards the island center. As this happens locally, we can analyze the mass transport and the different timescales involved in the reshaping of the island using SFM techniques.
To our knowledge, we here present the first experiment that, in addition to SFM measurements, simultaneously monitors the changes in the local contact potential of the WL via KPFM.
One of our central findings is that the Pb density of the WL 
decreases globally without showing an additional local depletion of Pb in a small neighborhood around the manipulated island.
Our results provide new insight into the nature of mass transport on Pb/Si(111)-($7\times7$).

\section{Experimental setup}
\label{sec:experiment}
We use phosphorus-doped n-type Si(111) ($1$ to $10$ $\Omega$cm) as a substrate, which we introduce into our ultra-high vacuum chamber with a base pressure of $\approx 10^{-10}~$mbar. The substrate is cleaned by heating it to a temperature of $700^\circ$C by direct current heating. The 7$\times$7-reconstructed Si(111) surface is created by rapidly increasing and decreasing the temperature to $1200^\circ$C and back to $700^\circ$C for at least five times and by a subsequent final cooling step, where the substrate temperature is kept between $850^\circ$C and $700^\circ$C for around $150$~s. 
We have prepared several substrates to grow Pb islands, where we first evaporate between 2 and 7 nominal monolayer (ML) Pb both on a substrate kept at room temperature and on a cooled substrate ($120$~K).
For the experiments using a cooled substrate during evaporation we then increase the sample temperature step-wise to $300$~K for imaging.
Our samples show consistent behavior.

For the SFM measurements we use a VT-SPM by \textit{Omicron}, with Si cantilevers by \textit{Nanosensors} with a force constant of $\approx 60~$Nm$^{-1}$ and a resonance frequency of $\approx 170~$kHz.
The tip is cleaned through several cycles of Ar-ion-sputtering and annealing.
Our experiment is conducted in FM-modulation mode. In parallel to the SFM topography measurement, the local contact potential difference (LCPD) is measured via KPFM using an AC bias applied through a lock-in with a modulation frequency of $440$~Hz and a modulation amplitude of $500$~mV.

To induce a local imbalance of Pb we enforce a contact between the island and the tip by decreasing the frequency shift to more negative values until we observe a strong additional decrease of the frequency shift, a sudden increase in the measured topography signal, see inset in Fig.~\ref{fig:kollision}~(b), and a slip of the phase locked loop phase regulation. For example for the experiment shown here, the set point of the frequency shift $\Delta f$ was slowly decreased from $-21$~Hz to $-47$~Hz.
Each image has an acquisition time of $31$~min and $t = 0$~min is set to the time of the first pixel measurement after the tip-island contact.
The slow scanning direction is the $y$-direction of the image, indicated in Fig~\ref{fig:kollision}~(e), while the $x$-direction corresponds to the fast scan direction. We show one out of four additional realizations of the island manipulation in Appendix~\ref{app:add_data}.

\section{Results}
\label{sec:results}
We focus on a sample with 5 ML Pb evaporated on a cooled Si substrate ($120$~K). 
After increasing the sample temperature to $300$~K, we observe large stable Pb islands (see Fig.~\ref{fig:kollision}~(a)). We first study the islands for $\approx 2$~h, where no growth in height is observed. Figs.~\ref{fig:kollision}~(a)-(e) show topographic measurements of the island before, during, and after the enforced tip-island contact. In Figs.~\ref{fig:kollision}~(f)-(j) the simultaneously obtained KPFM images show the measured potential difference.
The step edges of the Si substrate run parallel to the $y$-direction of the image, see Appendix~\ref{app:stepedges}. This causes an anisotropy of the substrate.
In the following discussion, we will mainly focus on islands I1, I2 and I3. 

Fig.~\ref{fig:kollision}~(b) shows the SFM measurement containing the moment of the enforced tip-sample contact. 
A horizontal scan line containing the spike marks the region in time where the collision occurred.
We attribute any changes that we observe subsequently to the short and local interaction between the tip and the island.
Fig.~\ref{fig:kollision}~(g) shows the LCPD measured in parallel to image Fig.~\ref{fig:kollision}~(b). 
The measured LCPD, both on the WL and on the island, remains unaltered after the contact. 
This demonstrates that through the tip-sample contact no material from the tip has been deposited onto the sample surface.
It also evidences that the opposite effect, i.e. Pb atoms contaminating or changing the tip geometry, is negligible. If this were the case, either the work function difference and thus the LCPD would show a strong shift immediately after contact. 
Or, if the geometry were to change, this would have an influence on the topographical measurement.
Neither of those effects is observed.
We therefore conclude that the contact does not change the tip in such a way that our measurement would be compromised.

Figs.~\ref{fig:kollision}~(b), (c), (g) and (h) show that also the neighboring islands have remained unchanged, i.e. with this method we solely manipulate the island locally and do not manipulate the WL or surrounding islands.

When we zoom out, we see that no alternations have been caused by scanning the sample.
No differences are found between the forward and backward scan, and the energy dissipation channel, i.e. the amplitude channel, shows no additional energy dissipation related to the tip-sample interaction during imaging.

The spike in the inset of Fig.~\ref{fig:kollision}~(b) shows that for a short time it was no longer possible to regulate the amplitude of the tip.
As a result the topographic image values are without any quantitative meaning for one pixel (7~ms).
This corresponds to the contact between the tip and the island surface.
By doing this we introduce additional energy to the system on a short timescale and in a strongly localized fashion. 
Meanwhile, no extra Pb is deposited.
Therefore, the increase or decrease of the island volume is directly connected to changes observed on the surface.

Already in the first scan line after the contact, we observe that a ring starts to form (see Fig.~\ref{fig:kollision}~(b)). In Fig.~\ref{fig:kollision}~(c) the ring on island I2 is fully closed and of equal height throughout. In Fig.~\ref{fig:kollision}~(e) the island top layer is fully closed and I2 assumes a hexagonal shape. We observe two processes that contribute to the filling of the ring of island I2: Firstly, the inner diameter decreases, such that the ring is contracted towards the island center. Secondly, filled layers form inside the ring. Throughout the reshaping of the island it grows in its radial dimension and obtains a fully hexagonal shape, which is the energetically favorable island shape (see Figs.~\ref{fig:kollision}~(c)-(e))~\cite{Liu_2013}.

In the topographic measurements (see Figs.~\ref{fig:kollision}~(a)-(e)) no change of the WL around island I2 is observed.
It is well-known that the Pb density of the WL is larger than the Pb density of crystalline Pb, i.e. the WL contains between $1.2$-$1.3$~ML Pb~\cite{hupalo01p1,Man08p1,Li09p1}. 
In general, the Pb density of the WL depends on the temperature of the system~\cite{hupalo01p1}. 
To investigate the Pb density of the surrounding WL and to answer the question what the source of the additional Pb in island I2 is, we examine the measured LCPD signal.
We expect that the LCPD is related to the chemical composition of the WL, i.e. to the density of Pb atoms in the WL.
This allows us to observe Pb density changes in the WL, which are not resolved in the SFM-topography measurement. 

Surprisingly, the KPFM measurements before, during and after the tip-island contact (Figs.~\ref{fig:kollision}~(f)-(j)) show that the WL density decreases uniformly in the whole measurement image. 
Additionally, between Fig.~\ref{fig:kollision}~(f) to Fig.~\ref{fig:kollision}~(j), we observe an increase of the mean LCPD $\Delta\overline{\Phi}$ of the WL from $94$~mV to $114$~mV as shown by the color shift of the ``background'' from dark orange to yellow.
For details see Appendix~\ref{app:meanLCPD}.

The color scale in Fig.~\ref{fig:kollision}~(f)-(j) emphasizes the changes in the WL.
Therefore, the quantum size effect (QSE) is difficult to recognize.
To show that the QSE is resolved in our measurement, we present the same data with a different choice of color scale in Appendix~\ref{app:QSE}.

In order to study the mass transport more quantitatively, we have additionally calculated the magnitude of the mass transport from the images.
This is done by normalizing the data by subtracting the mean value level and then summing over the measured island's pixel values.
The results are shown in Fig.~\ref{fig:CoM}~(a).
After the contact it takes 7~min 40~s to fully scan island I2 in Fig.~\ref{fig:kollision}~(b). In this time we observe a massive increase in the island volume of almost $30\%$, compared to island I2 in Fig.~\ref{fig:kollision}~(a).
This corresponds to a volumetric current of $\partial V \approx 0.29\cdot10^{3}~\mathrm{nm}^3\mathrm{/s}$ or $2360~\mathrm{atoms/s}$ (see Fig.~\ref{fig:CoM}~(a)).
This strong increase in the island volume raises the question: where does the additional Pb come from?

\section{Discussion}
\label{sec:discussion}
During imaging the tip interacts with the sample causing mechanical stress on the sample.
It is known that this force causes small distortions of the surface atoms, which can be reversible or irreversible~\cite{Lantz_2001}. 
In the experiment here presented all our observations point to a reversible motion, when the frequency shift is large, i.e. $-37$ Hz. 
The exception is the enforced tip-island contact with $\Delta f = - 47$ Hz, which is irreversible.
In the literature such irreversible atomic motions have been studied in the context of atomic jumps~\cite{Hoffmann_2007} and atomic manipulations~\cite{Custance_2009}. 

During the short and localized contact mechanical stress is applied to the island, which leads to its deformation and pushes Pb atoms onto the island top.
Directly after the contact more Pb material moves to the island top independent of the SFM tip.
It is reasonable to assume that this occurs to reach a new equilibrium island shape, which is strongly driven by the interplay between the classical step effect and the quantum size effect (QSE)~\cite{SLi06p01}. 
It has previously been shown that ring formation is the (energetically) favored growth mode for Pb islands~\cite{Li09p1,kuntov2008,spaeth20p1,kuntova2007}. The ring that is formed on top of island I2 adds another \blue{14} ML to the island height, as shown in Figs.~\ref{fig:kollision}~(b)-(c). This is consistent with the prediction of the QSE~\cite{spaeth17p1,spaeth20p1,Wei_2002,Jeffrey_2006,Schulte_1976,Bauer_88,Jian_2003,Su_2001}, i.e., a reversal of stable island heights (even-odd oscillation) for islands with heights larger than 10 ML. Hence, the new island height of 23 ML is a stable island height~\cite{spaeth20p1}. In comparable experiments, height changes usually range from one to two ML (e.g. see Ref.~\cite{spaeth20p1}). 
At temperatures above $T = 240$~K, multilayer (4 to 5 ML) ring growth has been observed~\cite{kuntova2007,Tringides2011_book}.

\begin{figure}
    \includegraphics[width=0.48\textwidth]{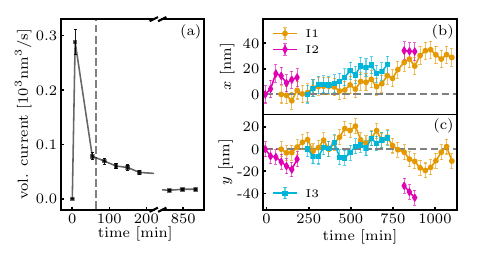}
    \caption{(a) Volumetric current flux of the incoming Pb atoms to island I2, showing a large increase in the island volume. The relative error is estimated to be $8\%$. (b) Change in the center of mass (COM) position of islands I1 (orange), I2 (magenta), and I3 (blue), see Fig.~\ref{fig:kollision}~(a). (b) All three islands' COM move vertically to the step edges (positive $x$-direction). Island I2 grows towards lower step edges, whereas island I1 and I3 decay from higher to lower step edges. 
    This leads to the same effective motion of the center of mass for all three islands. 
    (c) The $y$-components for island I1 and island I3 fluctuate around the origin, while island I2 shows an effective COM movement parallel to the Si step edges.
    The measurement time $t = 0$~min is set to the moment of the tip-island contact.}
    \label{fig:CoM}
\end{figure}

The start of ring formation is observed in the first scan line after the contact (Fig.~\ref{fig:kollision}~(b)). 
In Fig.~\ref{fig:kollision}~(c) the ring on island I2 is fully closed and of equal height throughout.
We find an upper bound of the ring formation time $\tau_{\mathrm{ring}} \leq 39$~min, given by the acquisition time of the measurement setup. 
In contrast, the ring closure towards the island center is orders of magnitude slower (i.e. $\tau_{\mathrm{radial}} \sim 1100$~min), Figs.~\ref{fig:kollision}~(d)-(e). 
In addition, after the initial ring formation, the volumetric current drops to $\partial V \sim 0.08\cdot10^{3}$ $\mathrm{nm}^3/\mathrm{s}$ (Fig. \ref{fig:CoM}~(a)). In Ref.~\cite{kuntova2007} the radial diffusion of Pb atoms towards the island center is assumed to be 1000 times slower than the azimuthal diffusion along the edge on top of the Pb island. This highly anisotropic diffusion behavior is also observed here for island I2, as $\tau_{\mathrm{ring}} \ll \tau_{\mathrm{radial}}$. 
Therefore, the manipulation of islands does trigger physical processes that occur on the same timescales as discussed in previous studies~\cite{Li09p1,kuntova2007}. 
This means that our method can explicitly be used to study these dynamics.

Our KPFM measurements allow us to study changes of the wetting layer around the manipulated island I2. 
The changes in the mean LCPD $\Delta\overline{\Phi}$ observed in a small neighborhood (see black dashed line in Fig.~\ref{fig:kollision}~(h)) around island I2 are similar to the changes observed in the small neighborhoods of island I1 and I3.
This is remarkable considering the changes in the shape and height of island I2 after the local tip-island contact. 
In fact, instead of finding a local depletion around island I2, we observe a ``global'' thinning of the WL throughout our measurement frame (where the ``global'' scale is indicated by the green dashed-dotted line in Fig.~\ref{fig:kollision}~(h)). 
As the number of Pb atoms in the system remains constant, the mass transfer to the island occurs solely via the WL.
From Fig.~\ref{fig:CoM}~(a), it is evident that the volume of the manipulated island increases as the volumetric current is positive ($\partial V > 0$) throughout the observation time. 

Due to the local tip-island contact, we do not directly modify the WL.
Changes in the WL are either caused by the transformation of island I2 after the tip-island contact or by the small increase in our sample temperature $\Delta T = 1.28$~K over the measurement time of $15$~h. 
It remains unclear how strongly the temperature change influences the WL density, however it is reasonable to assume that such small changes do not significantly alter the LCPD.

In contrast to other experiments, e.g. Ref.~\cite{Man08p1}, we here suggest a collective thinning of the WL. 
This means that the density of the Pb atoms in the WL appears to locally stay constant, similar to Ref.~\cite{hupalo07p1}.
Despite that, there is an effective mass transport happening, which causes the island growth.
As we have no ``free'' Pb atoms, due to our long annealing time, our experimental evidence points at a collective transport of Pb atoms from the WL towards island I2.

To rule out the possibility that this effect is caused by the tip, we contrast the changes $\Delta\overline{\Phi}$ in the mean LCPD of the WL with the islands, see Appendix~\ref{app:meanLCPD}.
We find that the island mean LCPD remains constant over the entire measurement period, hence indicating that the thinning of the WL is due to intrinsic processes on the surface.

Focusing on the neighboring islands I1 and I3 (see Figs.~\ref{fig:kollision}~(a) and (f)), we observe that they decay. 
Island I1 shows a decrease in volume from $V_1(t = 95~\mathrm{min}) = 334\cdot10^{3}~\mathrm{nm}^3$ to $V_1(t = 1040~\mathrm{min}) = 247\cdot10^{3}~\mathrm{nm}^3$ or by roughly $26\%$.
This decay can be seen by comparing island I1 in Fig.~\ref{fig:kollision}~(d) with Fig.~\ref{fig:kollision}~(e), where the side on the higher step edge fades away.
Similarly, island I3 decreases from $V_3(t = 126~\mathrm{min}) = 183\cdot10^{3}~\mathrm{nm}^3$ to $V_3(t = 630~\mathrm{min}) = 132\cdot10^{3}~\mathrm{nm}^3$ or by $28\%$.
The changes in the volume of these neighboring islands cannot account for the absolute volume increase of island I2, as more additional material is needed.
Additionally, we find that the decays of island I1 and I3 are time-delayed compared to the growing of island I2, e.g. the volume of I1 remains constant for about $10$ hours after the island manipulation, $V_1(t = 95~\mathrm{min}) \approx V_1(t = 591~\mathrm{min}) = 334\cdot10^{3}~\mathrm{nm}^3$.
We propose that the unbound Pb atoms of the decaying islands replenish the thinning WL.
In addition, we conjecture that the thinning of the WL results in a change of its chemical potential. 
This favors the decay of energetically less stable islands~\cite{Li_2007}. 
Other islands, outside of our measurement frame, may also contribute to the replenishing of the thinning WL.

To quantify the changes of the growing and decaying islands (I1 - I3), we calculate the movement of their center of mass.
For that, we need a drift-free reference frame, which we determine using fixed points on the Si substrate, i.e. step edges.
These fixed points are identified in each measurement image and their mean position becomes the center of origin of the new reference frame.
The relative changes in the center of mass positions are shown in Fig.~\ref{fig:CoM}~(b) and (c).

The manipulated island grows along the $x$-direction towards lower step edges and tends to grow parallel to the step edges along the negative $y$-direction. The decay of the energetically less stable islands  I1 and I3 results in a shift of the center of mass (COM) towards island I2 in the $x$-direction (see Fig.~\ref{fig:CoM}~(b)).
In contrast, the $y$-component of the COM of island I1 and island I3 shows no effective change, but only fluctuates around the initial $y$-coordinate (see Fig.~\ref{fig:CoM}~(c)).
The similarity between the COM motion of island I1 and island I3 might indicate a correlated movement of the Pb atoms to lower step edges in the WL due to the surface anisotropy~\cite{hershberger14p1}.

\section{Conclusion}
We successfully generated a local imbalance on a Pb island by enforcing a contact between measurement tip and island.
We find that this leads to the rapid formation of a ring on the manipulated island. 
We observe two characteristic timescales, one for the formation of the ring ($t \leq 39$~min) and one for the radial closure towards the island center and the radial growth of the island ($t > 1000$~min).
One key advantage of our local manipulation is the conservation of Pb atoms in the system. 

By simultaneously carrying out KPFM measurements, we are able to observe an increase of the wetting layer LCPD for some time after the tip contact.
This indicates a global thinning of the WL, instead of a local Pb depletion around the locally manipulated island.

Understanding the mechanisms (experimentally and theoretically) of this collective, global evolution of the WL is crucial to further characterize the diffusive mass transport of Pb/Si(111)-($7\times7$).
For example, it remains a task for future experiments to study the temperature dependence of the WL density.
Improved resolution will also allow us to use quantitative models and analysis
tools~\cite{Metzler_2014,Mu_oz_Gil_2021} to study in more detail the question of whether the motion
in the wetting layer is superdiffusive~\cite{Tringides2011_book} and/or correlated, as well as to
pinpoint the parameters of the island center-of-mass motion as function of
island size, temperature, and lifetime~\cite{hershberger14p1,Brune_1998,Evans_2006}.
The results will be also relevant to answer the outstanding question about the ``ultrafast'' initial island nucleation~\cite{jaroch2019,hershberger14p1}.

\section{Acknowledgments}
P.P.S. and F.H. thank Daniel Rothhardt for helpful discussions and comments on the subject of this research. 
This work is funded by the German Science Foundation (DFG Grant No. 513075417).
\bibliographystyle{apsrev4-2} 
\bibliography{main}

\begin{thebibliography}{44}%
\makeatletter
\providecommand \@ifxundefined [1]{%
 \@ifx{#1\undefined}
}%
\providecommand \@ifnum [1]{%
 \ifnum #1\expandafter \@firstoftwo
 \else \expandafter \@secondoftwo
 \fi
}%
\providecommand \@ifx [1]{%
 \ifx #1\expandafter \@firstoftwo
 \else \expandafter \@secondoftwo
 \fi
}%
\providecommand \natexlab [1]{#1}%
\providecommand \enquote  [1]{``#1''}%
\providecommand \bibnamefont  [1]{#1}%
\providecommand \bibfnamefont [1]{#1}%
\providecommand \citenamefont [1]{#1}%
\providecommand \href@noop [0]{\@secondoftwo}%
\providecommand \href [0]{\begingroup \@sanitize@url \@href}%
\providecommand \@href[1]{\@@startlink{#1}\@@href}%
\providecommand \@@href[1]{\endgroup#1\@@endlink}%
\providecommand \@sanitize@url [0]{\catcode `\\12\catcode `\$12\catcode
  `\&12\catcode `\#12\catcode `\^12\catcode `\_12\catcode `\%12\relax}%
\providecommand \@@startlink[1]{}%
\providecommand \@@endlink[0]{}%
\providecommand \url  [0]{\begingroup\@sanitize@url \@url }%
\providecommand \@url [1]{\endgroup\@href {#1}{\urlprefix }}%
\providecommand \urlprefix  [0]{URL }%
\providecommand \Eprint [0]{\href }%
\providecommand \doibase [0]{https://doi.org/}%
\providecommand \selectlanguage [0]{\@gobble}%
\providecommand \bibinfo  [0]{\@secondoftwo}%
\providecommand \bibfield  [0]{\@secondoftwo}%
\providecommand \translation [1]{[#1]}%
\providecommand \BibitemOpen [0]{}%
\providecommand \bibitemStop [0]{}%
\providecommand \bibitemNoStop [0]{.\EOS\space}%
\providecommand \EOS [0]{\spacefactor3000\relax}%
\providecommand \BibitemShut  [1]{\csname bibitem#1\endcsname}%
\let\auto@bib@innerbib\@empty
\bibitem [{\citenamefont {Venables}\ \emph {et~al.}(1984)\citenamefont
  {Venables}, \citenamefont {Spiller},\ and\ \citenamefont
  {Hanbucken}}]{Venables84p1}%
  \BibitemOpen
  \bibfield  {author} {\bibinfo {author} {\bibfnamefont {J.~A.}\ \bibnamefont
  {Venables}}, \bibinfo {author} {\bibfnamefont {G.~D.~T.}\ \bibnamefont
  {Spiller}},\ and\ \bibinfo {author} {\bibfnamefont {M.}~\bibnamefont
  {Hanbucken}},\ }\href {https://doi.org/10.1088%2F0034-4885%2F47%2F4%2F002}
  {\bibfield  {journal} {\bibinfo  {journal} {Rep. Prog. Phys.}\ }\textbf
  {\bibinfo {volume} {47}},\ \bibinfo {pages} {399} (\bibinfo {year}
  {1984})}\BibitemShut {NoStop}%
\bibitem [{\citenamefont {Baskaran}\ and\ \citenamefont
  {Smereka}(2012)}]{Baskaran_2012}%
  \BibitemOpen
  \bibfield  {author} {\bibinfo {author} {\bibfnamefont {A.}~\bibnamefont
  {Baskaran}}\ and\ \bibinfo {author} {\bibfnamefont {P.}~\bibnamefont
  {Smereka}},\ }\href {http://dx.doi.org/10.1063/1.3679068} {\bibfield
  {journal} {\bibinfo  {journal} {J. Appl. Phys.}\ }\textbf {\bibinfo {volume}
  {111}},\ \bibinfo {pages} {044321} (\bibinfo {year} {2012})}\BibitemShut
  {NoStop}%
\bibitem [{\citenamefont {Kesaria}\ \emph {et~al.}(2009)\citenamefont
  {Kesaria}, \citenamefont {Kumar}, \citenamefont {Govind},\ and\ \citenamefont
  {Shivaprasad}}]{Kesaria_2009}%
  \BibitemOpen
  \bibfield  {author} {\bibinfo {author} {\bibfnamefont {M.}~\bibnamefont
  {Kesaria}}, \bibinfo {author} {\bibfnamefont {M.}~\bibnamefont {Kumar}},
  \bibinfo {author} {\bibnamefont {Govind}},\ and\ \bibinfo {author}
  {\bibfnamefont {S.}~\bibnamefont {Shivaprasad}},\ }\href
  {https://www.sciencedirect.com/science/article/pii/S016943320901201X}
  {\bibfield  {journal} {\bibinfo  {journal} {Appl. Surf. Sci.}\ }\textbf
  {\bibinfo {volume} {256}},\ \bibinfo {pages} {576–579} (\bibinfo {year}
  {2009})}\BibitemShut {NoStop}%
\bibitem [{\citenamefont {Prieto}\ and\ \citenamefont
  {Markov}(2017)}]{Prieto_2017}%
  \BibitemOpen
  \bibfield  {author} {\bibinfo {author} {\bibfnamefont {J.}~\bibnamefont
  {Prieto}}\ and\ \bibinfo {author} {\bibfnamefont {I.}~\bibnamefont
  {Markov}},\ }\href {http://dx.doi.org/10.1016/j.susc.2017.05.018} {\bibfield
  {journal} {\bibinfo  {journal} {Surf. Sci.}\ }\textbf {\bibinfo {volume}
  {664}},\ \bibinfo {pages} {172–184} (\bibinfo {year} {2017})}\BibitemShut
  {NoStop}%
\bibitem [{\citenamefont {Bauer}(1958)}]{BAUER_1958}%
  \BibitemOpen
  \bibfield  {author} {\bibinfo {author} {\bibfnamefont {E.}~\bibnamefont
  {Bauer}},\ }\href {http://dx.doi.org/10.1524/zkri.1958.110.16.372} {\bibfield
   {journal} {\bibinfo  {journal} {Z. Kristallogr. – Cryst. Mater.}\ }\textbf
  {\bibinfo {volume} {110}},\ \bibinfo {pages} {372–394} (\bibinfo {year}
  {1958})}\BibitemShut {NoStop}%
\bibitem [{\citenamefont {Hershberger}\ \emph {et~al.}(2014)\citenamefont
  {Hershberger}, \citenamefont {Hupalo}, \citenamefont {Thiel}, \citenamefont
  {Wang}, \citenamefont {Ho},\ and\ \citenamefont
  {Tringides}}]{hershberger14p1}%
  \BibitemOpen
  \bibfield  {author} {\bibinfo {author} {\bibfnamefont {M.~T.}\ \bibnamefont
  {Hershberger}}, \bibinfo {author} {\bibfnamefont {M.}~\bibnamefont {Hupalo}},
  \bibinfo {author} {\bibfnamefont {P.~A.}\ \bibnamefont {Thiel}}, \bibinfo
  {author} {\bibfnamefont {C.~Z.}\ \bibnamefont {Wang}}, \bibinfo {author}
  {\bibfnamefont {K.~M.}\ \bibnamefont {Ho}},\ and\ \bibinfo {author}
  {\bibfnamefont {M.~C.}\ \bibnamefont {Tringides}},\ }\href
  {https://link.aps.org/doi/10.1103/PhysRevLett.113.236101} {\bibfield
  {journal} {\bibinfo  {journal} {Phys. Rev. Lett.}\ }\textbf {\bibinfo
  {volume} {113}},\ \bibinfo {pages} {236101} (\bibinfo {year}
  {2014})}\BibitemShut {NoStop}%
\bibitem [{\citenamefont {Hupalo}\ \emph {et~al.}(2001)\citenamefont {Hupalo},
  \citenamefont {Kremmer}, \citenamefont {Yeh}, \citenamefont
  {Berbil-Bautista}, \citenamefont {Abram},\ and\ \citenamefont
  {Tringides}}]{hupalo01p1}%
  \BibitemOpen
  \bibfield  {author} {\bibinfo {author} {\bibfnamefont {M.}~\bibnamefont
  {Hupalo}}, \bibinfo {author} {\bibfnamefont {S.}~\bibnamefont {Kremmer}},
  \bibinfo {author} {\bibfnamefont {V.}~\bibnamefont {Yeh}}, \bibinfo {author}
  {\bibfnamefont {L.}~\bibnamefont {Berbil-Bautista}}, \bibinfo {author}
  {\bibfnamefont {E.}~\bibnamefont {Abram}},\ and\ \bibinfo {author}
  {\bibfnamefont {M.}~\bibnamefont {Tringides}},\ }\href
  {https://doi.org/10.1016%2Fs0039-6028%2801%2901262-6} {\bibfield  {journal}
  {\bibinfo  {journal} {Surf. Sci.}\ }\textbf {\bibinfo {volume} {493}},\
  \bibinfo {pages} {526–538} (\bibinfo {year} {2001})}\BibitemShut {NoStop}%
\bibitem [{\citenamefont {Hong}\ \emph {et~al.}(2007)\citenamefont {Hong},
  \citenamefont {Basile}, \citenamefont {Czoschke}, \citenamefont {Gray},\ and\
  \citenamefont {Chiang}}]{Hong_2007}%
  \BibitemOpen
  \bibfield  {author} {\bibinfo {author} {\bibfnamefont {H.}~\bibnamefont
  {Hong}}, \bibinfo {author} {\bibfnamefont {L.}~\bibnamefont {Basile}},
  \bibinfo {author} {\bibfnamefont {P.}~\bibnamefont {Czoschke}}, \bibinfo
  {author} {\bibfnamefont {A.}~\bibnamefont {Gray}},\ and\ \bibinfo {author}
  {\bibfnamefont {T.-C.}\ \bibnamefont {Chiang}},\ }\href
  {http://dx.doi.org/10.1063/1.2435615} {\bibfield  {journal} {\bibinfo
  {journal} {Appl. Phys. Lett.}\ }\textbf {\bibinfo {volume} {90}},\ \bibinfo
  {pages} {051911} (\bibinfo {year} {2007})}\BibitemShut {NoStop}%
\bibitem [{\citenamefont {Jaroch}\ \emph {et~al.}(2019)\citenamefont {Jaroch},
  \citenamefont {Chen}, \citenamefont {Zdyb}, \citenamefont {Ja{\l}ochowski},
  \citenamefont {Thiel},\ and\ \citenamefont {Tringides}}]{jaroch2019}%
  \BibitemOpen
  \bibfield  {author} {\bibinfo {author} {\bibfnamefont {T.}~\bibnamefont
  {Jaroch}}, \bibinfo {author} {\bibfnamefont {S.}~\bibnamefont {Chen}},
  \bibinfo {author} {\bibfnamefont {R.}~\bibnamefont {Zdyb}}, \bibinfo {author}
  {\bibfnamefont {M.}~\bibnamefont {Ja{\l}ochowski}}, \bibinfo {author}
  {\bibfnamefont {P.}~\bibnamefont {Thiel}},\ and\ \bibinfo {author}
  {\bibfnamefont {M.}~\bibnamefont {Tringides}},\ }\href
  {https://doi.org/10.1016%2Fj.jcrysgro.2019.06.023} {\bibfield  {journal}
  {\bibinfo  {journal} {J. Cryst. Growth}\ }\textbf {\bibinfo {volume} {523}},\
  \bibinfo {pages} {125137} (\bibinfo {year} {2019})}\BibitemShut {NoStop}%
\bibitem [{\citenamefont {Ganz}\ \emph {et~al.}(1991)\citenamefont {Ganz},
  \citenamefont {Ing-Shouh}, \citenamefont {Fulin}, \citenamefont {Theiss},\
  and\ \citenamefont {Golovchenko}}]{Ganz_1991}%
  \BibitemOpen
  \bibfield  {author} {\bibinfo {author} {\bibfnamefont {E.}~\bibnamefont
  {Ganz}}, \bibinfo {author} {\bibfnamefont {H.}~\bibnamefont {Ing-Shouh}},
  \bibinfo {author} {\bibfnamefont {X.}~\bibnamefont {Fulin}}, \bibinfo
  {author} {\bibfnamefont {S.~K.}\ \bibnamefont {Theiss}},\ and\ \bibinfo
  {author} {\bibfnamefont {J.}~\bibnamefont {Golovchenko}},\ }\href
  {http://dx.doi.org/10.1016/0039-6028(91)90797-v} {\bibfield  {journal}
  {\bibinfo  {journal} {Surf. Sci.}\ }\textbf {\bibinfo {volume} {257}},\
  \bibinfo {pages} {259–273} (\bibinfo {year} {1991})}\BibitemShut {NoStop}%
\bibitem [{\citenamefont {Weitering}\ \emph {et~al.}(1992)\citenamefont
  {Weitering}, \citenamefont {Heslinga},\ and\ \citenamefont
  {Hibma}}]{Weitering_1992}%
  \BibitemOpen
  \bibfield  {author} {\bibinfo {author} {\bibfnamefont {H.~H.}\ \bibnamefont
  {Weitering}}, \bibinfo {author} {\bibfnamefont {D.~R.}\ \bibnamefont
  {Heslinga}},\ and\ \bibinfo {author} {\bibfnamefont {T.}~\bibnamefont
  {Hibma}},\ }\href {https://link.aps.org/doi/10.1103/PhysRevB.45.5991}
  {\bibfield  {journal} {\bibinfo  {journal} {Phys. Rev. B}\ }\textbf {\bibinfo
  {volume} {45}},\ \bibinfo {pages} {5991} (\bibinfo {year}
  {1992})}\BibitemShut {NoStop}%
\bibitem [{\citenamefont {Wang}\ \emph {et~al.}(2008)\citenamefont {Wang},
  \citenamefont {Zhang}, \citenamefont {Loy},\ and\ \citenamefont
  {Xiao}}]{Wang_2008}%
  \BibitemOpen
  \bibfield  {author} {\bibinfo {author} {\bibfnamefont {K.}~\bibnamefont
  {Wang}}, \bibinfo {author} {\bibfnamefont {X.}~\bibnamefont {Zhang}},
  \bibinfo {author} {\bibfnamefont {M.}~\bibnamefont {Loy}},\ and\ \bibinfo
  {author} {\bibfnamefont {X.}~\bibnamefont {Xiao}},\ }\href
  {http://dx.doi.org/10.1016/j.susc.2008.01.023} {\bibfield  {journal}
  {\bibinfo  {journal} {Surf. Sci.}\ }\textbf {\bibinfo {volume} {602}},\
  \bibinfo {pages} {1217–1222} (\bibinfo {year} {2008})}\BibitemShut
  {NoStop}%
\bibitem [{\citenamefont {Huang}\ \emph {et~al.}(2012)\citenamefont {Huang},
  \citenamefont {Wang}, \citenamefont {Li},\ and\ \citenamefont
  {Ho}}]{Huang2012}%
  \BibitemOpen
  \bibfield  {author} {\bibinfo {author} {\bibfnamefont {L.}~\bibnamefont
  {Huang}}, \bibinfo {author} {\bibfnamefont {C.~Z.}\ \bibnamefont {Wang}},
  \bibinfo {author} {\bibfnamefont {M.~Z.}\ \bibnamefont {Li}},\ and\ \bibinfo
  {author} {\bibfnamefont {K.~M.}\ \bibnamefont {Ho}},\ }\href
  {https://doi.org/10.1103%2Fphysrevlett.108.026101} {\bibfield  {journal}
  {\bibinfo  {journal} {Phys. Rev. Lett.}\ }\textbf {\bibinfo {volume} {108}},\
  \bibinfo {pages} {026101} (\bibinfo {year} {2012})}\BibitemShut {NoStop}%
\bibitem [{\citenamefont {Hupalo}\ and\ \citenamefont
  {Tringides}(2007)}]{hupalo07p1}%
  \BibitemOpen
  \bibfield  {author} {\bibinfo {author} {\bibfnamefont {M.}~\bibnamefont
  {Hupalo}}\ and\ \bibinfo {author} {\bibfnamefont {M.~C.}\ \bibnamefont
  {Tringides}},\ }\href {https://doi.org/10.1103/PhysRevB.75.235443} {\bibfield
   {journal} {\bibinfo  {journal} {Phys. Rev. B}\ }\textbf {\bibinfo {volume}
  {75}},\ \bibinfo {pages} {235443} (\bibinfo {year} {2007})}\BibitemShut
  {NoStop}%
\bibitem [{\citenamefont {Man}\ \emph {et~al.}(2008)\citenamefont {Man},
  \citenamefont {Tringides}, \citenamefont {Loy},\ and\ \citenamefont
  {Altman}}]{Man08p1}%
  \BibitemOpen
  \bibfield  {author} {\bibinfo {author} {\bibfnamefont {K.~L.}\ \bibnamefont
  {Man}}, \bibinfo {author} {\bibfnamefont {M.~C.}\ \bibnamefont {Tringides}},
  \bibinfo {author} {\bibfnamefont {M.~M.~T.}\ \bibnamefont {Loy}},\ and\
  \bibinfo {author} {\bibfnamefont {M.~S.}\ \bibnamefont {Altman}},\ }\href
  {https://link.aps.org/doi/10.1103/PhysRevLett.101.226102} {\bibfield
  {journal} {\bibinfo  {journal} {Phys. Rev. Lett.}\ }\textbf {\bibinfo
  {volume} {101}},\ \bibinfo {pages} {226102} (\bibinfo {year}
  {2008})}\BibitemShut {NoStop}%
\bibitem [{\citenamefont {Man}\ \emph {et~al.}(2013)\citenamefont {Man},
  \citenamefont {Tringides}, \citenamefont {Loy},\ and\ \citenamefont
  {Altman}}]{man2013}%
  \BibitemOpen
  \bibfield  {author} {\bibinfo {author} {\bibfnamefont {K.~L.}\ \bibnamefont
  {Man}}, \bibinfo {author} {\bibfnamefont {M.~C.}\ \bibnamefont {Tringides}},
  \bibinfo {author} {\bibfnamefont {M.~M.~T.}\ \bibnamefont {Loy}},\ and\
  \bibinfo {author} {\bibfnamefont {M.~S.}\ \bibnamefont {Altman}},\ }\href
  {https://doi.org/10.1103%2Fphysrevlett.110.036104} {\bibfield  {journal}
  {\bibinfo  {journal} {Phys. Rev. Lett.}\ }\textbf {\bibinfo {volume} {110}},\
  \bibinfo {pages} {036104} (\bibinfo {year} {2013})}\BibitemShut {NoStop}%
\bibitem [{\citenamefont {Kuntov{\'{a}}}\ \emph {et~al.}(2008)\citenamefont
  {Kuntov{\'{a}}}, \citenamefont {Chvoj}, \citenamefont {Tringides},\ and\
  \citenamefont {Yakes}}]{kuntov2008}%
  \BibitemOpen
  \bibfield  {author} {\bibinfo {author} {\bibfnamefont {Z.}~\bibnamefont
  {Kuntov{\'{a}}}}, \bibinfo {author} {\bibfnamefont {Z.}~\bibnamefont
  {Chvoj}}, \bibinfo {author} {\bibfnamefont {M.~C.}\ \bibnamefont
  {Tringides}},\ and\ \bibinfo {author} {\bibfnamefont {M.}~\bibnamefont
  {Yakes}},\ }\href {https://doi.org/10.1140%2Fepjb%2Fe2008-00278-6} {\bibfield
   {journal} {\bibinfo  {journal} {Eur. Phys. J. B}\ }\textbf {\bibinfo
  {volume} {64}},\ \bibinfo {pages} {61–66} (\bibinfo {year}
  {2008})}\BibitemShut {NoStop}%
\bibitem [{\citenamefont {Li}\ \emph {et~al.}(2009)\citenamefont {Li},
  \citenamefont {Wang}, \citenamefont {Evans}, \citenamefont {Hupalo},
  \citenamefont {Tringides},\ and\ \citenamefont {Ho}}]{Li09p1}%
  \BibitemOpen
  \bibfield  {author} {\bibinfo {author} {\bibfnamefont {M.}~\bibnamefont
  {Li}}, \bibinfo {author} {\bibfnamefont {C.~Z.}\ \bibnamefont {Wang}},
  \bibinfo {author} {\bibfnamefont {J.~W.}\ \bibnamefont {Evans}}, \bibinfo
  {author} {\bibfnamefont {M.}~\bibnamefont {Hupalo}}, \bibinfo {author}
  {\bibfnamefont {M.~C.}\ \bibnamefont {Tringides}},\ and\ \bibinfo {author}
  {\bibfnamefont {K.~M.}\ \bibnamefont {Ho}},\ }\href
  {https://doi.org/10.1103%2Fphysrevb.79.113404} {\bibfield  {journal}
  {\bibinfo  {journal} {Phys. Rev. B}\ }\textbf {\bibinfo {volume} {79}},\
  \bibinfo {pages} {113404} (\bibinfo {year} {2009})}\BibitemShut {NoStop}%
\bibitem [{\citenamefont {Kuntova}\ \emph {et~al.}(2007)\citenamefont
  {Kuntova}, \citenamefont {Hupalo}, \citenamefont {Chvoj},\ and\ \citenamefont
  {Tringides}}]{kuntova2007}%
  \BibitemOpen
  \bibfield  {author} {\bibinfo {author} {\bibfnamefont {Z.}~\bibnamefont
  {Kuntova}}, \bibinfo {author} {\bibfnamefont {M.}~\bibnamefont {Hupalo}},
  \bibinfo {author} {\bibfnamefont {Z.}~\bibnamefont {Chvoj}},\ and\ \bibinfo
  {author} {\bibfnamefont {M.~C.}\ \bibnamefont {Tringides}},\ }\href
  {https://link.aps.org/doi/10.1103/PhysRevB.75.205436} {\bibfield  {journal}
  {\bibinfo  {journal} {Phys. Rev. B}\ }\textbf {\bibinfo {volume} {75}},\
  \bibinfo {pages} {205436} (\bibinfo {year} {2007})}\BibitemShut {NoStop}%
\bibitem [{\citenamefont {Kuntova}\ \emph {et~al.}(2006)\citenamefont
  {Kuntova}, \citenamefont {Hupalo}, \citenamefont {Chvoj},\ and\ \citenamefont
  {Tringides}}]{Kuntova2006}%
  \BibitemOpen
  \bibfield  {author} {\bibinfo {author} {\bibfnamefont {Z.}~\bibnamefont
  {Kuntova}}, \bibinfo {author} {\bibfnamefont {M.}~\bibnamefont {Hupalo}},
  \bibinfo {author} {\bibfnamefont {Z.}~\bibnamefont {Chvoj}},\ and\ \bibinfo
  {author} {\bibfnamefont {M.}~\bibnamefont {Tringides}},\ }\href
  {http://dx.doi.org/10.1016/j.susc.2006.07.052} {\bibfield  {journal}
  {\bibinfo  {journal} {Surf. Sci.}\ }\textbf {\bibinfo {volume} {600}},\
  \bibinfo {pages} {4765–4770} (\bibinfo {year} {2006})}\BibitemShut
  {NoStop}%
\bibitem [{\citenamefont {Tajik}\ \emph {et~al.}(2023)\citenamefont {Tajik},
  \citenamefont {Witte}, \citenamefont {Brand}, \citenamefont {Rettig},
  \citenamefont {Sothmann}, \citenamefont {Bovensiepen},\ and\ \citenamefont
  {von Hoegen}}]{Tajik_2023}%
  \BibitemOpen
  \bibfield  {author} {\bibinfo {author} {\bibfnamefont {M.}~\bibnamefont
  {Tajik}}, \bibinfo {author} {\bibfnamefont {T.}~\bibnamefont {Witte}},
  \bibinfo {author} {\bibfnamefont {C.}~\bibnamefont {Brand}}, \bibinfo
  {author} {\bibfnamefont {L.}~\bibnamefont {Rettig}}, \bibinfo {author}
  {\bibfnamefont {B.}~\bibnamefont {Sothmann}}, \bibinfo {author}
  {\bibfnamefont {U.}~\bibnamefont {Bovensiepen}},\ and\ \bibinfo {author}
  {\bibfnamefont {M.~H.}\ \bibnamefont {von Hoegen}},\ }\href
  {https://arxiv.org/abs/2312.04541} {\bibinfo {title} {Electron phonon
  coupling in ultrathin pb films on si(111): Where the heck is the energy?}}
  (\bibinfo {year} {2023}),\ \Eprint {https://arxiv.org/abs/2312.04541}
  {arXiv:2312.04541 [cond-mat.mes-hall]} \BibitemShut {NoStop}%
\bibitem [{\citenamefont {Kratzer}\ \emph {et~al.}(2022)\citenamefont
  {Kratzer}, \citenamefont {Rettig}, \citenamefont {Sklyadneva}, \citenamefont
  {Chulkov},\ and\ \citenamefont {Bovensiepen}}]{Kratzer_2022}%
  \BibitemOpen
  \bibfield  {author} {\bibinfo {author} {\bibfnamefont {P.}~\bibnamefont
  {Kratzer}}, \bibinfo {author} {\bibfnamefont {L.}~\bibnamefont {Rettig}},
  \bibinfo {author} {\bibfnamefont {I.~Y.}\ \bibnamefont {Sklyadneva}},
  \bibinfo {author} {\bibfnamefont {E.~V.}\ \bibnamefont {Chulkov}},\ and\
  \bibinfo {author} {\bibfnamefont {U.}~\bibnamefont {Bovensiepen}},\ }\href
  {http://dx.doi.org/10.1103/physrevresearch.4.033218} {\bibfield  {journal}
  {\bibinfo  {journal} {Phys. Rev. Research}\ }\textbf {\bibinfo {volume}
  {4}},\ \bibinfo {pages} {033218} (\bibinfo {year} {2022})}\BibitemShut
  {NoStop}%
\bibitem [{\citenamefont {Tresca}\ \emph {et~al.}(2023)\citenamefont {Tresca},
  \citenamefont {Bilgeri}, \citenamefont {Ménard}, \citenamefont {Cherkez},
  \citenamefont {Federicci}, \citenamefont {Longo}, \citenamefont {Hervé},
  \citenamefont {Debontridder}, \citenamefont {David}, \citenamefont
  {Roditchev}, \citenamefont {Profeta}, \citenamefont {Cren}, \citenamefont
  {Calandra},\ and\ \citenamefont {Brun}}]{Tresca_2023}%
  \BibitemOpen
  \bibfield  {author} {\bibinfo {author} {\bibfnamefont {C.}~\bibnamefont
  {Tresca}}, \bibinfo {author} {\bibfnamefont {T.}~\bibnamefont {Bilgeri}},
  \bibinfo {author} {\bibfnamefont {G.}~\bibnamefont {Ménard}}, \bibinfo
  {author} {\bibfnamefont {V.}~\bibnamefont {Cherkez}}, \bibinfo {author}
  {\bibfnamefont {R.}~\bibnamefont {Federicci}}, \bibinfo {author}
  {\bibfnamefont {D.}~\bibnamefont {Longo}}, \bibinfo {author} {\bibfnamefont
  {M.}~\bibnamefont {Hervé}}, \bibinfo {author} {\bibfnamefont
  {F.}~\bibnamefont {Debontridder}}, \bibinfo {author} {\bibfnamefont
  {P.}~\bibnamefont {David}}, \bibinfo {author} {\bibfnamefont
  {D.}~\bibnamefont {Roditchev}}, \bibinfo {author} {\bibfnamefont
  {G.}~\bibnamefont {Profeta}}, \bibinfo {author} {\bibfnamefont
  {T.}~\bibnamefont {Cren}}, \bibinfo {author} {\bibfnamefont {M.}~\bibnamefont
  {Calandra}},\ and\ \bibinfo {author} {\bibfnamefont {C.}~\bibnamefont
  {Brun}},\ }\href {http://dx.doi.org/10.1103/physrevb.107.035125} {\bibfield
  {journal} {\bibinfo  {journal} {Phys. Rev. B}\ }\textbf {\bibinfo {volume}
  {107}},\ \bibinfo {pages} {035125} (\bibinfo {year} {2023})}\BibitemShut
  {NoStop}%
\bibitem [{\citenamefont {Vandelli}\ \emph {et~al.}(2024)\citenamefont
  {Vandelli}, \citenamefont {Galler}, \citenamefont {Rubio}, \citenamefont
  {Lichtenstein}, \citenamefont {Biermann},\ and\ \citenamefont
  {Stepanov}}]{Vandelli_2024}%
  \BibitemOpen
  \bibfield  {author} {\bibinfo {author} {\bibfnamefont {M.}~\bibnamefont
  {Vandelli}}, \bibinfo {author} {\bibfnamefont {A.}~\bibnamefont {Galler}},
  \bibinfo {author} {\bibfnamefont {A.}~\bibnamefont {Rubio}}, \bibinfo
  {author} {\bibfnamefont {A.~I.}\ \bibnamefont {Lichtenstein}}, \bibinfo
  {author} {\bibfnamefont {S.}~\bibnamefont {Biermann}},\ and\ \bibinfo
  {author} {\bibfnamefont {E.~A.}\ \bibnamefont {Stepanov}},\ }\href
  {http://dx.doi.org/10.1038/s41535-024-00630-w} {\bibfield  {journal}
  {\bibinfo  {journal} {npj Quantum Mater.}\ }\textbf {\bibinfo {volume} {9}}
  (\bibinfo {year} {2024})}\BibitemShut {NoStop}%
\bibitem [{\citenamefont {Li}\ \emph {et~al.}(2006)\citenamefont {Li},
  \citenamefont {Ma}, \citenamefont {Jia}, \citenamefont {Zhang}, \citenamefont
  {Chen}, \citenamefont {Niu}, \citenamefont {Liu}, \citenamefont {Weiss},\
  and\ \citenamefont {Xue}}]{SLi06p01}%
  \BibitemOpen
  \bibfield  {author} {\bibinfo {author} {\bibfnamefont {S.-C.}\ \bibnamefont
  {Li}}, \bibinfo {author} {\bibfnamefont {X.}~\bibnamefont {Ma}}, \bibinfo
  {author} {\bibfnamefont {J.-F.}\ \bibnamefont {Jia}}, \bibinfo {author}
  {\bibfnamefont {Y.-F.}\ \bibnamefont {Zhang}}, \bibinfo {author}
  {\bibfnamefont {D.}~\bibnamefont {Chen}}, \bibinfo {author} {\bibfnamefont
  {Q.}~\bibnamefont {Niu}}, \bibinfo {author} {\bibfnamefont {F.}~\bibnamefont
  {Liu}}, \bibinfo {author} {\bibfnamefont {P.~S.}\ \bibnamefont {Weiss}},\
  and\ \bibinfo {author} {\bibfnamefont {Q.-K.}\ \bibnamefont {Xue}},\ }\href
  {https://doi.org/10.1103%2Fphysrevb.74.075410} {\bibfield  {journal}
  {\bibinfo  {journal} {Phys. Rev. B}\ }\textbf {\bibinfo {volume} {74}},\
  \bibinfo {pages} {195428} (\bibinfo {year} {2006})}\BibitemShut {NoStop}%
\bibitem [{\citenamefont {Han}\ \emph {et~al.}(2004)\citenamefont {Han},
  \citenamefont {Zhu}, \citenamefont {Liu}, \citenamefont {Li}, \citenamefont
  {Jia}, \citenamefont {Zhang},\ and\ \citenamefont {Xue}}]{Han04p1}%
  \BibitemOpen
  \bibfield  {author} {\bibinfo {author} {\bibfnamefont {Y.}~\bibnamefont
  {Han}}, \bibinfo {author} {\bibfnamefont {J.~Y.}\ \bibnamefont {Zhu}},
  \bibinfo {author} {\bibfnamefont {F.}~\bibnamefont {Liu}}, \bibinfo {author}
  {\bibfnamefont {S.-C.}\ \bibnamefont {Li}}, \bibinfo {author} {\bibfnamefont
  {J.-F.}\ \bibnamefont {Jia}}, \bibinfo {author} {\bibfnamefont {Y.-F.}\
  \bibnamefont {Zhang}},\ and\ \bibinfo {author} {\bibfnamefont {Q.-K.}\
  \bibnamefont {Xue}},\ }\href
  {https://link.aps.org/doi/10.1103/PhysRevLett.93.106102} {\bibfield
  {journal} {\bibinfo  {journal} {Phys. Rev. Lett.}\ }\textbf {\bibinfo
  {volume} {93}},\ \bibinfo {pages} {106102} (\bibinfo {year}
  {2004})}\BibitemShut {NoStop}%
\bibitem [{\citenamefont {Liu}\ \emph {et~al.}(2013)\citenamefont {Liu},
  \citenamefont {Xiao}, \citenamefont {Yang}, \citenamefont {Zhang},
  \citenamefont {Jiang}, \citenamefont {Fei}, \citenamefont {Du},\ and\
  \citenamefont {Gao}}]{Liu_2013}%
  \BibitemOpen
  \bibfield  {author} {\bibinfo {author} {\bibfnamefont {L.~W.}\ \bibnamefont
  {Liu}}, \bibinfo {author} {\bibfnamefont {W.~D.}\ \bibnamefont {Xiao}},
  \bibinfo {author} {\bibfnamefont {K.}~\bibnamefont {Yang}}, \bibinfo {author}
  {\bibfnamefont {L.~Z.}\ \bibnamefont {Zhang}}, \bibinfo {author}
  {\bibfnamefont {Y.~H.}\ \bibnamefont {Jiang}}, \bibinfo {author}
  {\bibfnamefont {X.~M.}\ \bibnamefont {Fei}}, \bibinfo {author} {\bibfnamefont
  {S.~X.}\ \bibnamefont {Du}},\ and\ \bibinfo {author} {\bibfnamefont {H.-J.}\
  \bibnamefont {Gao}},\ }\href {http://dx.doi.org/10.1021/jp404190c} {\bibfield
   {journal} {\bibinfo  {journal} {J. Phys. Chem. C}\ }\textbf {\bibinfo
  {volume} {117}},\ \bibinfo {pages} {22652–22655} (\bibinfo {year}
  {2013})}\BibitemShut {NoStop}%
\bibitem [{\citenamefont {Lantz}\ \emph {et~al.}(2001)\citenamefont {Lantz},
  \citenamefont {Hug}, \citenamefont {Hoffmann}, \citenamefont {van Schendel},
  \citenamefont {Kappenberger}, \citenamefont {Martin}, \citenamefont
  {Baratoff},\ and\ \citenamefont {Güntherodt}}]{Lantz_2001}%
  \BibitemOpen
  \bibfield  {author} {\bibinfo {author} {\bibfnamefont {M.~A.}\ \bibnamefont
  {Lantz}}, \bibinfo {author} {\bibfnamefont {H.~J.}\ \bibnamefont {Hug}},
  \bibinfo {author} {\bibfnamefont {R.}~\bibnamefont {Hoffmann}}, \bibinfo
  {author} {\bibfnamefont {P.~J.~A.}\ \bibnamefont {van Schendel}}, \bibinfo
  {author} {\bibfnamefont {P.}~\bibnamefont {Kappenberger}}, \bibinfo {author}
  {\bibfnamefont {S.}~\bibnamefont {Martin}}, \bibinfo {author} {\bibfnamefont
  {A.}~\bibnamefont {Baratoff}},\ and\ \bibinfo {author} {\bibfnamefont
  {H.-J.}\ \bibnamefont {Güntherodt}},\ }\href
  {http://dx.doi.org/10.1126/science.1057824} {\bibfield  {journal} {\bibinfo
  {journal} {Science}\ }\textbf {\bibinfo {volume} {291}},\ \bibinfo {pages}
  {2580–2583} (\bibinfo {year} {2001})}\BibitemShut {NoStop}%
\bibitem [{\citenamefont {Hoffmann}\ \emph {et~al.}(2007)\citenamefont
  {Hoffmann}, \citenamefont {Baratoff}, \citenamefont {Hug}, \citenamefont
  {Hidber}, \citenamefont {Löhneysen},\ and\ \citenamefont
  {Güntherodt}}]{Hoffmann_2007}%
  \BibitemOpen
  \bibfield  {author} {\bibinfo {author} {\bibfnamefont {R.}~\bibnamefont
  {Hoffmann}}, \bibinfo {author} {\bibfnamefont {A.}~\bibnamefont {Baratoff}},
  \bibinfo {author} {\bibfnamefont {H.~J.}\ \bibnamefont {Hug}}, \bibinfo
  {author} {\bibfnamefont {H.~R.}\ \bibnamefont {Hidber}}, \bibinfo {author}
  {\bibfnamefont {H.~v.}\ \bibnamefont {Löhneysen}},\ and\ \bibinfo {author}
  {\bibfnamefont {H.-J.}\ \bibnamefont {Güntherodt}},\ }\href
  {http://dx.doi.org/10.1088/0957-4484/18/39/395503} {\bibfield  {journal}
  {\bibinfo  {journal} {Nanotechnology}\ }\textbf {\bibinfo {volume} {18}},\
  \bibinfo {pages} {395503} (\bibinfo {year} {2007})}\BibitemShut {NoStop}%
\bibitem [{\citenamefont {Custance}\ \emph {et~al.}(2009)\citenamefont
  {Custance}, \citenamefont {Perez},\ and\ \citenamefont
  {Morita}}]{Custance_2009}%
  \BibitemOpen
  \bibfield  {author} {\bibinfo {author} {\bibfnamefont {O.}~\bibnamefont
  {Custance}}, \bibinfo {author} {\bibfnamefont {R.}~\bibnamefont {Perez}},\
  and\ \bibinfo {author} {\bibfnamefont {S.}~\bibnamefont {Morita}},\ }\href
  {http://dx.doi.org/10.1038/nnano.2009.347} {\bibfield  {journal} {\bibinfo
  {journal} {Nat. Nanotechnol.}\ }\textbf {\bibinfo {volume} {4}},\ \bibinfo
  {pages} {803–810} (\bibinfo {year} {2009})}\BibitemShut {NoStop}%
\bibitem [{\citenamefont {Sp\"{a}th}\ \emph {et~al.}(2020)\citenamefont
  {Sp\"{a}th}, \citenamefont {Popp},\ and\ \citenamefont
  {Hoffmann-Vogel}}]{spaeth20p1}%
  \BibitemOpen
  \bibfield  {author} {\bibinfo {author} {\bibfnamefont {T.}~\bibnamefont
  {Sp\"{a}th}}, \bibinfo {author} {\bibfnamefont {M.}~\bibnamefont {Popp}},\
  and\ \bibinfo {author} {\bibfnamefont {R.}~\bibnamefont {Hoffmann-Vogel}},\
  }\href {https://link.aps.org/doi/10.1103/PhysRevLett.124.016101} {\bibfield
  {journal} {\bibinfo  {journal} {Phys. Rev. Lett.}\ }\textbf {\bibinfo
  {volume} {124}},\ \bibinfo {pages} {016101} (\bibinfo {year}
  {2020})}\BibitemShut {NoStop}%
\bibitem [{\citenamefont {Sp\"{a}th}\ \emph {et~al.}(2017)\citenamefont
  {Sp\"{a}th}, \citenamefont {Popp}, \citenamefont {{P{\'{e}}rez Le{\'{o}}n}},
  \citenamefont {Marz},\ and\ \citenamefont {Hoffmann-Vogel}}]{spaeth17p1}%
  \BibitemOpen
  \bibfield  {author} {\bibinfo {author} {\bibfnamefont {T.}~\bibnamefont
  {Sp\"{a}th}}, \bibinfo {author} {\bibfnamefont {M.}~\bibnamefont {Popp}},
  \bibinfo {author} {\bibfnamefont {C.}~\bibnamefont {{P{\'{e}}rez
  Le{\'{o}}n}}}, \bibinfo {author} {\bibfnamefont {M.}~\bibnamefont {Marz}},\
  and\ \bibinfo {author} {\bibfnamefont {R.}~\bibnamefont {Hoffmann-Vogel}},\
  }\href {https://doi.org/10.1039/c7nr01874f} {\bibfield  {journal} {\bibinfo
  {journal} {Nanoscale}\ }\textbf {\bibinfo {volume} {9}},\ \bibinfo {pages}
  {7868–7874} (\bibinfo {year} {2017})}\BibitemShut {NoStop}%
\bibitem [{\citenamefont {Wei}\ and\ \citenamefont {Chou}(2002)}]{Wei_2002}%
  \BibitemOpen
  \bibfield  {author} {\bibinfo {author} {\bibfnamefont {C.~M.}\ \bibnamefont
  {Wei}}\ and\ \bibinfo {author} {\bibfnamefont {M.~Y.}\ \bibnamefont {Chou}},\
  }\href {https://doi.org/10.1103%2Fphysrevb.66.233408} {\bibfield  {journal}
  {\bibinfo  {journal} {Phys. Rev. B}\ }\textbf {\bibinfo {volume} {66}},\
  \bibinfo {pages} {233408} (\bibinfo {year} {2002})}\BibitemShut {NoStop}%
\bibitem [{\citenamefont {Jeffrey}\ \emph {et~al.}(2006)\citenamefont
  {Jeffrey}, \citenamefont {Conrad}, \citenamefont {Feng}, \citenamefont
  {Hupalo}, \citenamefont {Kim}, \citenamefont {Ryan}, \citenamefont {Miceli},\
  and\ \citenamefont {Tringides}}]{Jeffrey_2006}%
  \BibitemOpen
  \bibfield  {author} {\bibinfo {author} {\bibfnamefont {C.~A.}\ \bibnamefont
  {Jeffrey}}, \bibinfo {author} {\bibfnamefont {E.~H.}\ \bibnamefont {Conrad}},
  \bibinfo {author} {\bibfnamefont {R.}~\bibnamefont {Feng}}, \bibinfo {author}
  {\bibfnamefont {M.}~\bibnamefont {Hupalo}}, \bibinfo {author} {\bibfnamefont
  {C.}~\bibnamefont {Kim}}, \bibinfo {author} {\bibfnamefont {P.~J.}\
  \bibnamefont {Ryan}}, \bibinfo {author} {\bibfnamefont {P.~F.}\ \bibnamefont
  {Miceli}},\ and\ \bibinfo {author} {\bibfnamefont {M.~C.}\ \bibnamefont
  {Tringides}},\ }\href {https://doi.org/10.1103%2Fphysrevlett.96.106105}
  {\bibfield  {journal} {\bibinfo  {journal} {Phys. Rev. Lett.}\ }\textbf
  {\bibinfo {volume} {96}},\ \bibinfo {pages} {106105} (\bibinfo {year}
  {2006})}\BibitemShut {NoStop}%
\bibitem [{\citenamefont {Schulte}(1976)}]{Schulte_1976}%
  \BibitemOpen
  \bibfield  {author} {\bibinfo {author} {\bibfnamefont {F.}~\bibnamefont
  {Schulte}},\ }\href {http://dx.doi.org/10.1016/0039-6028(76)90250-8}
  {\bibfield  {journal} {\bibinfo  {journal} {Surf. Sci.}\ }\textbf {\bibinfo
  {volume} {55}},\ \bibinfo {pages} {427–444} (\bibinfo {year}
  {1976})}\BibitemShut {NoStop}%
\bibitem [{\citenamefont {Ja\l{}ochowski}\ and\ \citenamefont
  {Bauer}(1988)}]{Bauer_88}%
  \BibitemOpen
  \bibfield  {author} {\bibinfo {author} {\bibfnamefont {M.}~\bibnamefont
  {Ja\l{}ochowski}}\ and\ \bibinfo {author} {\bibfnamefont {E.}~\bibnamefont
  {Bauer}},\ }\href {https://link.aps.org/doi/10.1103/PhysRevB.38.5272}
  {\bibfield  {journal} {\bibinfo  {journal} {Phys. Rev. B}\ }\textbf {\bibinfo
  {volume} {38}},\ \bibinfo {pages} {5272} (\bibinfo {year}
  {1988})}\BibitemShut {NoStop}%
\bibitem [{\citenamefont {Jian}\ \emph {et~al.}(2003)\citenamefont {Jian},
  \citenamefont {Su}, \citenamefont {Chang},\ and\ \citenamefont
  {Tsong}}]{Jian_2003}%
  \BibitemOpen
  \bibfield  {author} {\bibinfo {author} {\bibfnamefont {W.~B.}\ \bibnamefont
  {Jian}}, \bibinfo {author} {\bibfnamefont {W.~B.}\ \bibnamefont {Su}},
  \bibinfo {author} {\bibfnamefont {C.~S.}\ \bibnamefont {Chang}},\ and\
  \bibinfo {author} {\bibfnamefont {T.~T.}\ \bibnamefont {Tsong}},\ }\href
  {https://link.aps.org/doi/10.1103/PhysRevLett.90.196603} {\bibfield
  {journal} {\bibinfo  {journal} {Phys. Rev. Lett.}\ }\textbf {\bibinfo
  {volume} {90}},\ \bibinfo {pages} {196603} (\bibinfo {year}
  {2003})}\BibitemShut {NoStop}%
\bibitem [{\citenamefont {Su}\ \emph {et~al.}(2001)\citenamefont {Su},
  \citenamefont {Chang}, \citenamefont {Chang}, \citenamefont {Chen},\ and\
  \citenamefont {Tsong}}]{Su_2001}%
  \BibitemOpen
  \bibfield  {author} {\bibinfo {author} {\bibfnamefont {W.-B.}\ \bibnamefont
  {Su}}, \bibinfo {author} {\bibfnamefont {S.-H.}\ \bibnamefont {Chang}},
  \bibinfo {author} {\bibfnamefont {C.-S.}\ \bibnamefont {Chang}}, \bibinfo
  {author} {\bibfnamefont {L.~J.}\ \bibnamefont {Chen}},\ and\ \bibinfo
  {author} {\bibfnamefont {T.~T.}\ \bibnamefont {Tsong}},\ }\href
  {http://dx.doi.org/10.1143/jjap.40.4299} {\bibfield  {journal} {\bibinfo
  {journal} {Jpn. J. Appl. Phys.}\ }\textbf {\bibinfo {volume} {40}},\ \bibinfo
  {pages} {4299} (\bibinfo {year} {2001})}\BibitemShut {NoStop}%
\bibitem [{\citenamefont {Tringides}\ \emph {et~al.}(2011)\citenamefont
  {Tringides}, \citenamefont {Hupalo}, \citenamefont {Man}, \citenamefont
  {Loy},\ and\ \citenamefont {Altman}}]{Tringides2011_book}%
  \BibitemOpen
  \bibfield  {author} {\bibinfo {author} {\bibfnamefont {M.}~\bibnamefont
  {Tringides}}, \bibinfo {author} {\bibfnamefont {M.}~\bibnamefont {Hupalo}},
  \bibinfo {author} {\bibfnamefont {K.}~\bibnamefont {Man}}, \bibinfo {author}
  {\bibfnamefont {M.}~\bibnamefont {Loy}},\ and\ \bibinfo {author}
  {\bibfnamefont {M.}~\bibnamefont {Altman}},\ }\bibinfo {title} {Wetting
  $\mathrm{Layer}$ $\mathrm{Super}$-$\mathrm{Diffusive}$ $\mathrm{Motion}$ and
  $\mathrm{QSE}$ $\mathrm{Growth}$ in $\mathrm{Pb}/\mathrm{Si}$},\ in\ \href
  {https://doi.org/10.1007/978-3-642-16510-8_3} {\emph {\bibinfo {booktitle}
  {Nanophenomena at Surfaces: Fundamentals of Exotic Condensed Matter
  Properties}}},\ \bibinfo {editor} {edited by\ \bibinfo {editor}
  {\bibfnamefont {M.}~\bibnamefont {Michailov}}}\ (\bibinfo  {publisher}
  {Springer Berlin Heidelberg},\ \bibinfo {address} {Berlin, Heidelberg},\
  \bibinfo {year} {2011})\ pp.\ \bibinfo {pages} {39--65}\BibitemShut {NoStop}%
\bibitem [{\citenamefont {Li}\ \emph {et~al.}(2007)\citenamefont {Li},
  \citenamefont {Evans}, \citenamefont {Wang}, \citenamefont {Hupalo},
  \citenamefont {Tringides}, \citenamefont {Chan},\ and\ \citenamefont
  {Ho}}]{Li_2007}%
  \BibitemOpen
  \bibfield  {author} {\bibinfo {author} {\bibfnamefont {M.}~\bibnamefont
  {Li}}, \bibinfo {author} {\bibfnamefont {J.}~\bibnamefont {Evans}}, \bibinfo
  {author} {\bibfnamefont {C.}~\bibnamefont {Wang}}, \bibinfo {author}
  {\bibfnamefont {M.}~\bibnamefont {Hupalo}}, \bibinfo {author} {\bibfnamefont
  {M.}~\bibnamefont {Tringides}}, \bibinfo {author} {\bibfnamefont {T.-L.}\
  \bibnamefont {Chan}},\ and\ \bibinfo {author} {\bibfnamefont
  {K.}~\bibnamefont {Ho}},\ }\href
  {http://dx.doi.org/10.1016/j.susc.2007.08.029} {\bibfield  {journal}
  {\bibinfo  {journal} {Surf. Sci.}\ }\textbf {\bibinfo {volume} {601}},\
  \bibinfo {pages} {L140–L144} (\bibinfo {year} {2007})}\BibitemShut
  {NoStop}%
\bibitem [{\citenamefont {Metzler}\ \emph {et~al.}(2014)\citenamefont
  {Metzler}, \citenamefont {Jeon}, \citenamefont {Cherstvy},\ and\
  \citenamefont {Barkai}}]{Metzler_2014}%
  \BibitemOpen
  \bibfield  {author} {\bibinfo {author} {\bibfnamefont {R.}~\bibnamefont
  {Metzler}}, \bibinfo {author} {\bibfnamefont {J.-H.}\ \bibnamefont {Jeon}},
  \bibinfo {author} {\bibfnamefont {A.~G.}\ \bibnamefont {Cherstvy}},\ and\
  \bibinfo {author} {\bibfnamefont {E.}~\bibnamefont {Barkai}},\ }\href
  {http://dx.doi.org/10.1039/c4cp03465a} {\bibfield  {journal} {\bibinfo
  {journal} {Phys. Chem. Chem. Phys.}\ }\textbf {\bibinfo {volume} {16}},\
  \bibinfo {pages} {24128–24164} (\bibinfo {year} {2014})}\BibitemShut
  {NoStop}%
\bibitem [{\citenamefont {Muñoz-Gil}\ \emph {et~al.}(2021)\citenamefont
  {Muñoz-Gil}, \citenamefont {Volpe}, \citenamefont {Garcia-March},
  \citenamefont {Aghion}, \citenamefont {Argun}, \citenamefont {Hong},
  \citenamefont {Bland}, \citenamefont {Bo}, \citenamefont {Conejero},
  \citenamefont {Firbas}, \citenamefont {Garibo~i Orts}, \citenamefont
  {Gentili}, \citenamefont {Huang}, \citenamefont {Jeon}, \citenamefont
  {Kabbech}, \citenamefont {Kim}, \citenamefont {Kowalek}, \citenamefont
  {Krapf}, \citenamefont {Loch-Olszewska}, \citenamefont {Lomholt},
  \citenamefont {Masson}, \citenamefont {Meyer}, \citenamefont {Park},
  \citenamefont {Requena}, \citenamefont {Smal}, \citenamefont {Song},
  \citenamefont {Szwabiński}, \citenamefont {Thapa}, \citenamefont {Verdier},
  \citenamefont {Volpe}, \citenamefont {Widera}, \citenamefont {Lewenstein},
  \citenamefont {Metzler},\ and\ \citenamefont {Manzo}}]{Mu_oz_Gil_2021}%
  \BibitemOpen
  \bibfield  {author} {\bibinfo {author} {\bibfnamefont {G.}~\bibnamefont
  {Muñoz-Gil}}, \bibinfo {author} {\bibfnamefont {G.}~\bibnamefont {Volpe}},
  \bibinfo {author} {\bibfnamefont {M.~A.}\ \bibnamefont {Garcia-March}},
  \bibinfo {author} {\bibfnamefont {E.}~\bibnamefont {Aghion}}, \bibinfo
  {author} {\bibfnamefont {A.}~\bibnamefont {Argun}}, \bibinfo {author}
  {\bibfnamefont {C.~B.}\ \bibnamefont {Hong}}, \bibinfo {author}
  {\bibfnamefont {T.}~\bibnamefont {Bland}}, \bibinfo {author} {\bibfnamefont
  {S.}~\bibnamefont {Bo}}, \bibinfo {author} {\bibfnamefont {J.~A.}\
  \bibnamefont {Conejero}}, \bibinfo {author} {\bibfnamefont {N.}~\bibnamefont
  {Firbas}}, \bibinfo {author} {\bibfnamefont {{\`O}.}~\bibnamefont {Garibo~i
  Orts}}, \bibinfo {author} {\bibfnamefont {A.}~\bibnamefont {Gentili}},
  \bibinfo {author} {\bibfnamefont {Z.}~\bibnamefont {Huang}}, \bibinfo
  {author} {\bibfnamefont {J.-H.}\ \bibnamefont {Jeon}}, \bibinfo {author}
  {\bibfnamefont {H.}~\bibnamefont {Kabbech}}, \bibinfo {author} {\bibfnamefont
  {Y.}~\bibnamefont {Kim}}, \bibinfo {author} {\bibfnamefont {P.}~\bibnamefont
  {Kowalek}}, \bibinfo {author} {\bibfnamefont {D.}~\bibnamefont {Krapf}},
  \bibinfo {author} {\bibfnamefont {H.}~\bibnamefont {Loch-Olszewska}},
  \bibinfo {author} {\bibfnamefont {M.~A.}\ \bibnamefont {Lomholt}}, \bibinfo
  {author} {\bibfnamefont {J.-B.}\ \bibnamefont {Masson}}, \bibinfo {author}
  {\bibfnamefont {P.~G.}\ \bibnamefont {Meyer}}, \bibinfo {author}
  {\bibfnamefont {S.}~\bibnamefont {Park}}, \bibinfo {author} {\bibfnamefont
  {B.}~\bibnamefont {Requena}}, \bibinfo {author} {\bibfnamefont
  {I.}~\bibnamefont {Smal}}, \bibinfo {author} {\bibfnamefont {T.}~\bibnamefont
  {Song}}, \bibinfo {author} {\bibfnamefont {J.}~\bibnamefont {Szwabiński}},
  \bibinfo {author} {\bibfnamefont {S.}~\bibnamefont {Thapa}}, \bibinfo
  {author} {\bibfnamefont {H.}~\bibnamefont {Verdier}}, \bibinfo {author}
  {\bibfnamefont {G.}~\bibnamefont {Volpe}}, \bibinfo {author} {\bibfnamefont
  {A.}~\bibnamefont {Widera}}, \bibinfo {author} {\bibfnamefont
  {M.}~\bibnamefont {Lewenstein}}, \bibinfo {author} {\bibfnamefont
  {R.}~\bibnamefont {Metzler}},\ and\ \bibinfo {author} {\bibfnamefont
  {C.}~\bibnamefont {Manzo}},\ }\href
  {http://dx.doi.org/10.1038/s41467-021-26320-w} {\bibfield  {journal}
  {\bibinfo  {journal} {Nature Comm.}\ }\textbf {\bibinfo {volume} {12}},\
  \bibinfo {pages} {6253} (\bibinfo {year} {2021})}\BibitemShut {NoStop}%
\bibitem [{\citenamefont {Brune}(1998)}]{Brune_1998}%
  \BibitemOpen
  \bibfield  {author} {\bibinfo {author} {\bibfnamefont {H.}~\bibnamefont
  {Brune}},\ }\href {http://dx.doi.org/10.1016/s0167-5729(99)80001-6}
  {\bibfield  {journal} {\bibinfo  {journal} {Surf. Sci. Rep.}\ }\textbf
  {\bibinfo {volume} {31}},\ \bibinfo {pages} {125–229} (\bibinfo {year}
  {1998})}\BibitemShut {NoStop}%
\bibitem [{\citenamefont {Evans}\ \emph {et~al.}(2006)\citenamefont {Evans},
  \citenamefont {Thiel},\ and\ \citenamefont {Bartelt}}]{Evans_2006}%
  \BibitemOpen
  \bibfield  {author} {\bibinfo {author} {\bibfnamefont {J.}~\bibnamefont
  {Evans}}, \bibinfo {author} {\bibfnamefont {P.}~\bibnamefont {Thiel}},\ and\
  \bibinfo {author} {\bibfnamefont {M.}~\bibnamefont {Bartelt}},\ }\href
  {http://dx.doi.org/10.1016/j.surfrep.2005.08.004} {\bibfield  {journal}
  {\bibinfo  {journal} {Surf. Sci. Rep.}\ }\textbf {\bibinfo {volume} {61}},\
  \bibinfo {pages} {1–128} (\bibinfo {year} {2006})}\BibitemShut {NoStop}%
\end{thebibliography}%
\appendix
\section{Step edges of the Si surface}
\label{app:stepedges}
\begin{figure}[h]
    \centering
    \includegraphics[width=0.45\textwidth]{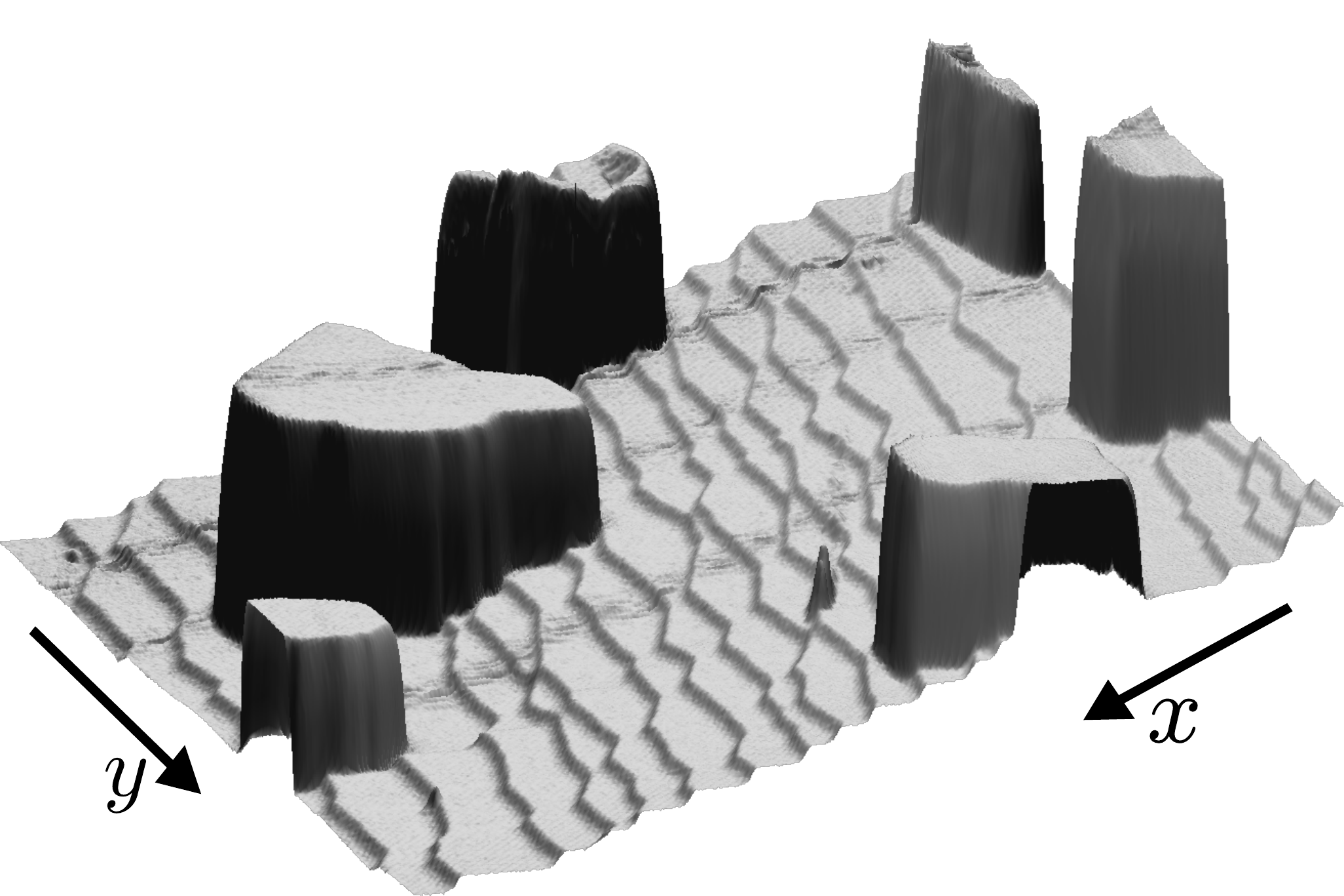}
    \caption{Here we show a 3D representation of Fig.~\ref{fig:kollision}~(a), with focus on the Si-substrate. The $y$-direction is defined parallel to the step edges and the $x$-direction downwards the step edges. For comparison see coordinate frame in Fig.~\ref{fig:kollision}~(e) of the main text.}
    \label{fig:stepedges}
\end{figure}

\section{Height-profiles of island I2}
\label{app:heightprofile}
\begin{figure*}
    \centering
    \includegraphics[width=\textwidth]{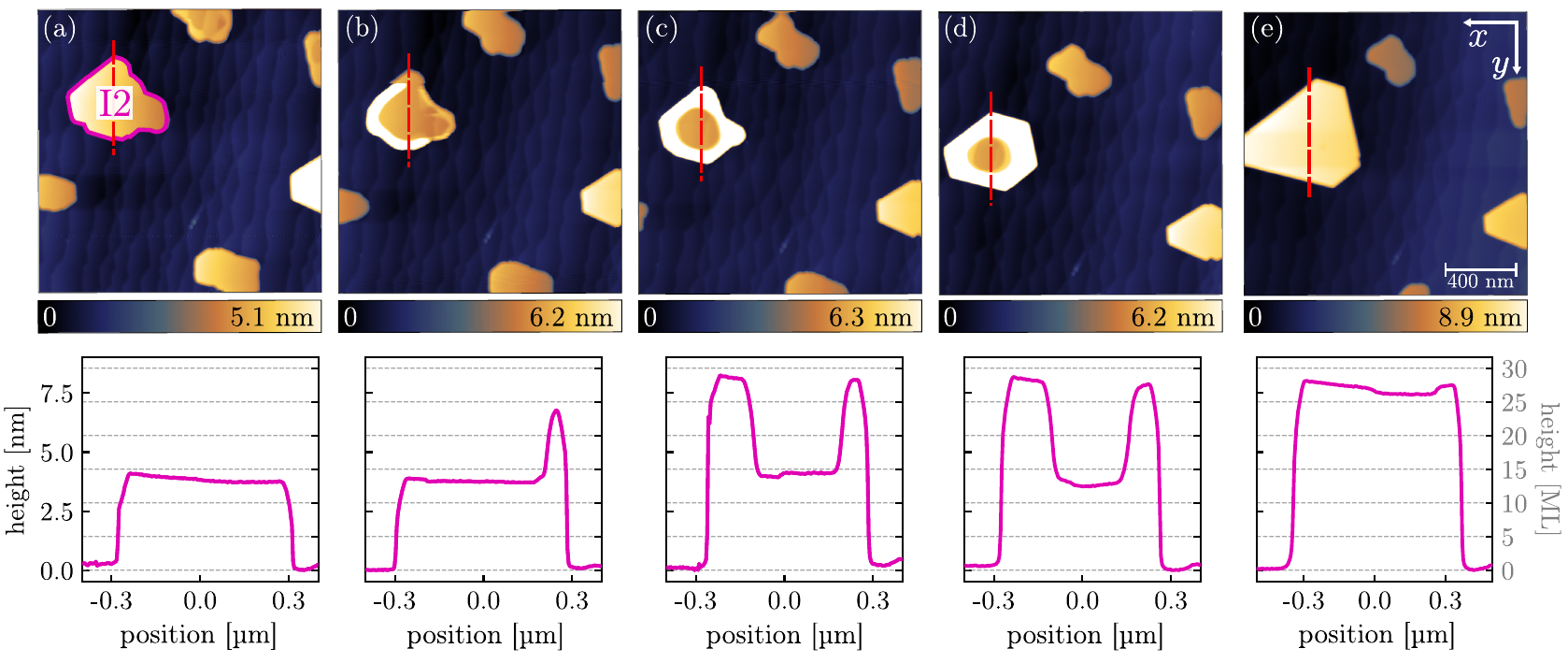}
    \caption{The same SFM images of 5 ML Pb on Si(111) as in Fig.~\ref{fig:kollision}~(a)-(e).
    Here the plane-leveling is such that the Si substrate has the same height throughout the image and the different island heights are empasized. 
    The lower row shows the height profiles of island I2 measured along the $y$-direction indicated by the red dashed line (parallel to the step edges).
    The $y$-axis on the left measures the island height in nanometers, and the $y$-axis on the right shows the island height in monolayers of bulk Pb(111) (emphasized by the gray dashed lines in the background).
    The formation of the ring, its closure towards the island center and the increase in the total island volume is clearly observed.
    Measurement details are given in Fig.~\ref{fig:kollision}.}
    \label{fig:heightprofiles}
\end{figure*}

In Fig.~\ref{fig:heightprofiles} we show the same SFM-topography data as in Fig.~\ref{fig:kollision}~(a)-(e), but here levelled with respect to the Si substrate. 

Moreover, in the lower row of Fig.~\ref{fig:heightprofiles}, we plot the height profiles of the manipulated island I2. The position of the cross-section of each plot is indicated by the red-dashed line in the SFM images above. 
The profiles themselves are centered around the island midpoint. 
The $y$-axis on the left depicts the island height measured in nanometers above the Si substrate. The $y$-axis on the right, together with the dashed gray lines, indicate the island height in monolayers of bulk Pb(111), i.e. one monolayer corresponds to a height of $0.285$~nm.

\section{Quantum size effect (QSE)}
\label{app:QSE}
\begin{figure*}
    \centering
    \includegraphics[width=\textwidth]{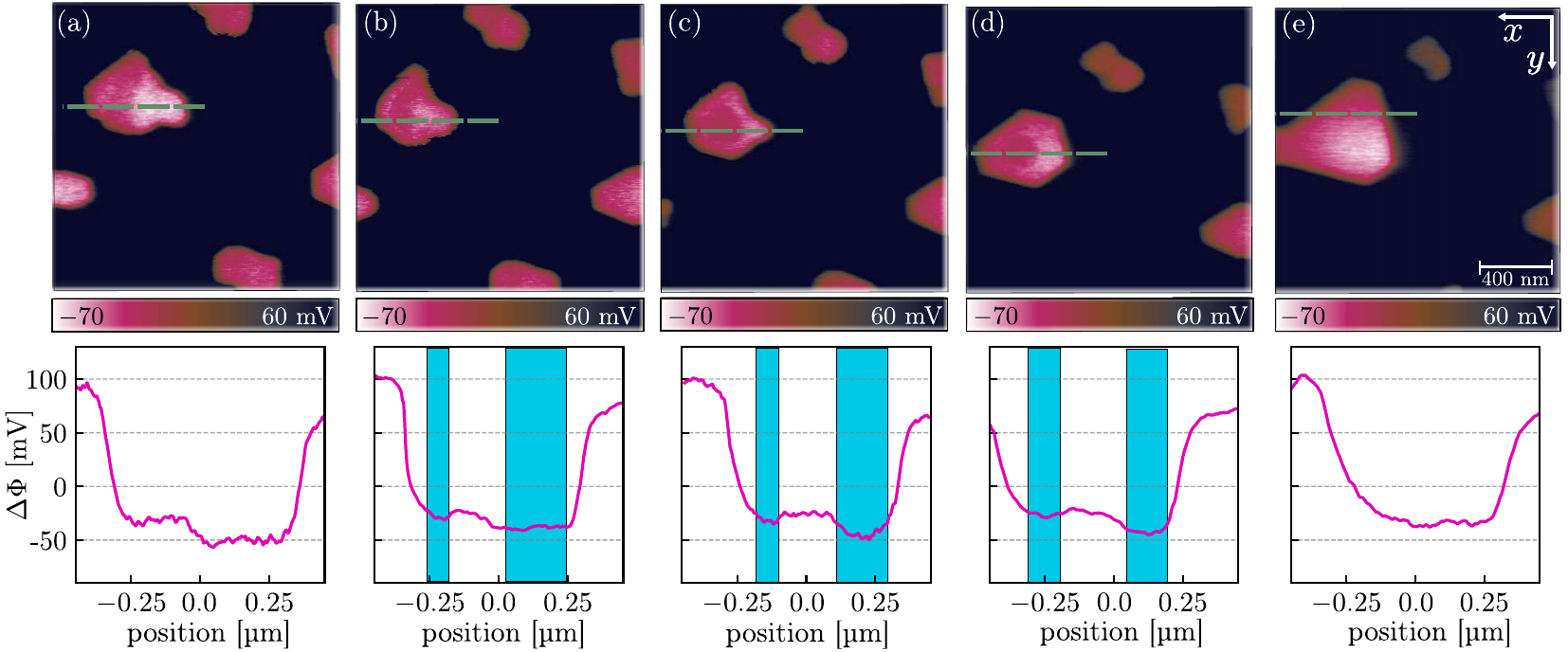}
    \caption{The presented LCPD data is identical to the data shown in Fig.~\ref{fig:kollision}~(f)-(j) with a different color scale. 
    Here we have adjusted the color scale such that the focus is on the QSE, instead of focusing on the changes in the WL. In (a)-(d) the QSE is clearly visible, by the measured differences of the LCPD on the island top, i.e. the different colors on the same island top. 
    Line profiles of the measured LCPD are shown in the second row. Their locations are indicated by the green dashed lines above. The light blue background indicates the position of the ring in the profiles. Compared to the LCPD of the island center, the LCPD measured at the ring is lower, indicating a stable island height.
    }
    \label{fig:QSE}
\end{figure*}

In Fig.~\ref{fig:QSE} we show the same LCPD data as in Fig.~\ref{fig:kollision}~(f)-(j), but we adjusted the color scale such that the QSE is amplified.
The QSE leads to an oscillation of the measured LCPD of the Pb electrons with an increasing number of Pb monolayers of the island.
This is due to the confinement of the Pb electrons between the vacuum and the Si substrate along the direction perpendicular to the sample surface.
The ratio between the Fermi wavelength of the Pb electrons ($\lambda_{\mathrm{F}}=0.394$~nm) and the thickness of a ML of PB ($d_0 = 0.285$~nm) is given by $2 d_0 \approx 1.5 \lambda_{\mathrm{F}}$.
This results in a bilayer oscillation of stable (low LCPD) and unstable (greater LCPD) island heights, i.e. odd heights are stable and even ones are unstable. 
This behavior is reversed after around 10 to 12 ML, due to the slight mismatch of the ratio between the Fermi wavelength and the ML thickness of Pb.
In addition, for increasing island heights the difference between the measured LCPD of odd and even island heights decreases.

The QSE creates a contrast (between white and red areas) observed on the islands in Fig.~\ref{fig:QSE}. 
In Fig.~\ref{fig:QSE}(a), the steps under the substrate ensure a different number of Pb monolayers within island I2, which results in a changing white-red contrast along the $x$-axis.
This is emphasized by the line profile along the green-dashed line plotted below Fig.~\ref{fig:QSE}(a), where the differences in the LCPD are observed. 
Though the island spans six terraces, only two distinct areas are observed. 
This is due to a double step under the island and spatially narrow terraces near the edges of the island. 

With the creation of the ring three distinct areas become visible (in the LCPD data and in the corresponding line profiles, see Fig.~\ref{fig:QSE}(b)-(d)).
The positions of the developing ring are highlighted with blue color in the line profiles. 
The island height (including the ring), on the left of island I2, is $28$~ML (left blue bar in the line profiles). 
The inner hole has a higher LCPD, as shown in the line profiles and by the red center area on island I2 (see Fig.~\ref{fig:QSE}(b)-(d)). 
The ring on the right edge of the island show again a lower LCPD (second blue bar in the line profiles). 

In Fig.~\ref{fig:QSE}~(e) the inner ring has fully vanished and the whole island has reached a flat top. The island consists of areas between $23$~ML to $28$~ML, at which the QSE becomes weaker and weaker.
\section{Mean LCPD of the wetting layer and the islands}
\label{app:meanLCPD}
\begin{figure}[b]
    \centering
    \includegraphics[width=0.49\textwidth]{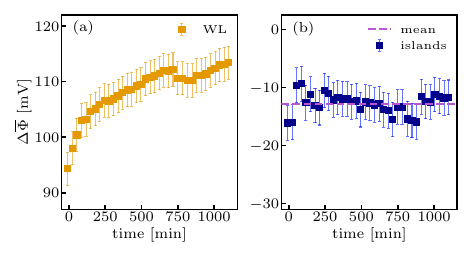}
    \caption{(a) Mean LCPD $\Delta\overline{\Phi}$ over the WL (yellow squares) and (b) the islands (blue squares). 
    We observe that $\Delta\overline{\Phi}$ of the WL increases over time.
    In contrast, the $\Delta\overline{\Phi}$ of the islands stays constant, with the mean of the data-points indicated by the purple-dashed line.
    The drop of $\Delta\overline{\Phi}$ at around $750$~min is due to a tip artifact.
    The error bars indicate the systematic error of our measurement setup, i.e. $\approx 3$~mV.
    }
    \label{fig:meanLCPD}
\end{figure}

In order to show that there is an effective change of the WL density, we compute the mean LCPD $\Delta\bar{\Phi}$ over the WL (yellow squares) and over the islands (blue squares) for each measurement frame, see Fig.~\ref{fig:meanLCPD}.
Throughout the measurement we observe an increase of $\Delta\bar{\Phi}$ from $94$~mV to $114$~mV. 
In contrast, we observe a constant mean LCPD over the islands, which shows larger fluctuations due to the oscillatory behavior of the LCPD caused by the QSE.
The averaged $\Delta\bar{\Phi}$ over time is indicated by the purple-dashed line (Fig.~\ref{fig:meanLCPD}~(b)).

This implies that the changes observed in the WL are not caused by the tip, which would lead to the same changes in the WL and on the islands. 
As mentioned in the main text, we conclude that the density of the WL decreases and provides Pb atoms for the reshaping of the manipulated island I2.

As $\Delta\bar{\Phi}$ is an averaged quantity over $N_{\mathrm{islands}} \approx 30000$~px and $N_{\mathrm{WL}} \approx 220000$~px, the statistical error of the mean is of the order of $\approx 0.2$~mV and $\approx0.03$~mV, respectively. The systematic error of our measurement apparatus is $\approx 3$~mV, which is indicated by the error bars in Fig.~\ref{fig:meanLCPD}.

\section{Additional experimental data}
\label{app:add_data}
\begin{figure*}
    \centering
    \includegraphics[width=0.9\textwidth]{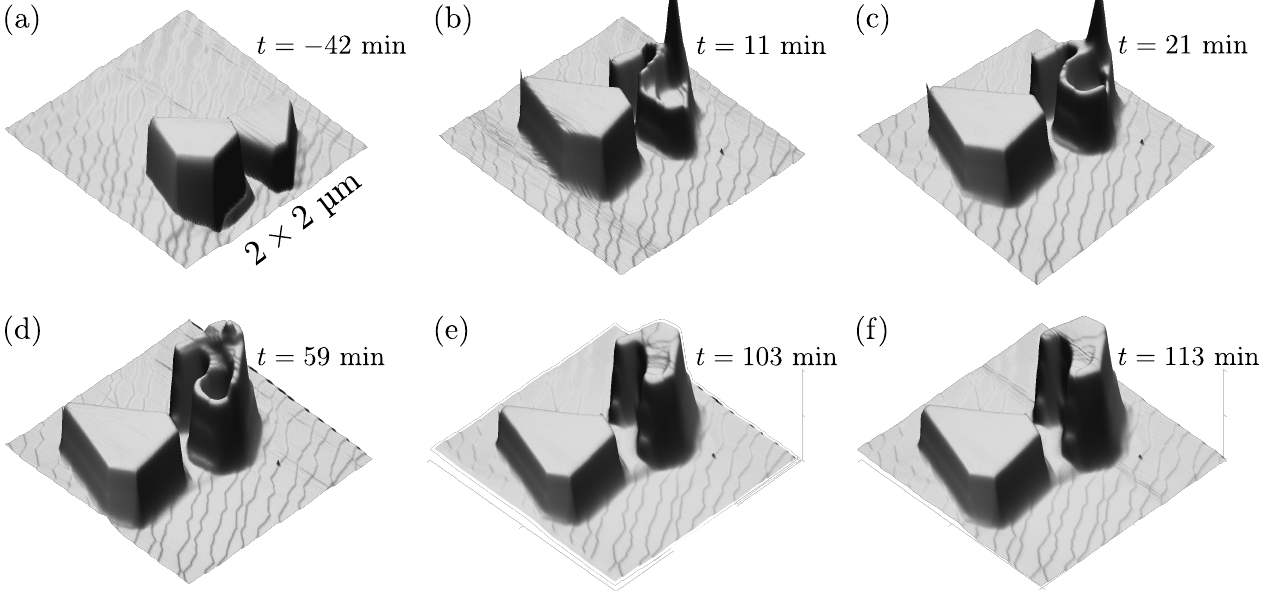}
    \caption{SFM measurement and manipulation by SFM tip of $2$~ML Pb on Si(111), which was deposited at a substrate temperature of $120$~K. (a) The measurement is performed after an annealing time of $19$~h, during which the sample equilibrated to room temperature, i.e. $296$~K. (b) Compared to the experiment discussed in the main text, the enforced tip-island contact is much stronger. The formation of a ring is again observed right after the contact. (c)-(f) Similar as discussed above, the ring first forms and then closes towards its center. In the end, the manipulated island reaches a new equilibrium state at a stable island height of $19$~nm. Measurement settings are $\Delta f=-37$~Hz with an oscillation amplitude $A_{\mathrm{osc}} = 7.9$~nm, an eigenfrequency of $f_0 = 155.81$~kHz, and a spring constant of $c_{\mathrm{L}} = 41$~N/m.}
    \label{fig:add_data}
\end{figure*}

In Fig.~\ref{fig:add_data}, we present an additional experiment in which we have enforced a contact between the cantilever and a stable island (right island in Fig.~\ref{fig:add_data}~(a)).
We evaporated $2$~ML of Pb on Si(111) at $120$~K. Subsequently, the cooling was stopped and the sample equilibrated to room temperature, i.e. $296$~K, over an annealing time of $19$~h.

Similar to the experiment discussed in the main text, right after the contact the island starts to grow in height by the formation of ring(s), see Fig.~\ref{fig:add_data}~(b).
Over the time of the measurement the ring closes toward its center (Fig.~\ref{fig:add_data}~(c) and (d)) and the island reaches a new stable configuration (see Fig.~\ref{fig:add_data}~(e) and (f)).

We emphasize that the observed physical mechanisms are comparable, even though the amount of evaporated Pb, the annealing time and the strength of the tip-island contact are different. 
This indicates that our experimental technique is able to reproduce the above discussed results and can be used to purposefully manipulate Pb islands.

\end{document}